\documentclass[11pt]{article}
 
\usepackage[a4paper,margin=1in]{geometry}
\usepackage[T1]{fontenc}
\usepackage[utf8]{inputenc}
\usepackage[title]{appendix}
 
\usepackage{amsmath,amsfonts}
\usepackage{mathtools}
\usepackage{algorithmic}
\usepackage{algorithm}
\usepackage{array}
\usepackage{indentfirst}
\usepackage[caption=false,font=normalsize,labelfont=sf,textfont=sf]{subfig}
\usepackage{textcomp}
\usepackage{xcolor}
\usepackage{url}
\usepackage{verbatim}
\usepackage{graphicx}
\usepackage{cite}
\usepackage{authblk}
 
\makeatletter
\renewcommand{\thesubsubsection}{\arabic{subsubsection}}
\renewcommand{\subsubsection}[1]{%
  \refstepcounter{subsubsection}
  \textbf{\thesubsubsection. #1}
}
\makeatother

\newlength{\abstractwidth}
\newenvironment{indexterms}
  {\begin{center}\normalsize\textbf{Index Terms}\\[0.5em]
   \begin{minipage}{\abstractwidth}\centering\ignorespaces}
  {\end{minipage}\end{center}}

\newcommand{\IEEEmembership}[1]{\textit{#1}}

\newcommand{\IEEEPARstart}[2]{#1#2}

\usepackage{footmisc}
\makeatletter
\let\oldthanks\thanks
\renewcommand{\thanks}[1]{\begingroup\def\@makefnmark{}\oldthanks{#1}\endgroup}
\makeatother

\makeatletter
\renewcommand{\thanks}[1]{%
  \footnotemark[0]%
  \protected@xdef\@thanks{\@thanks%
    \protect\footnotetext[0]{#1}}%
}
\makeatother

\begin{document}

\title{On Channel Model to Bridge the Gap between MIMO Design and Performance Requirements in 3GPP}

\author{Lynda~Berrah,~\IEEEmembership{Student Member,~IEEE,}
Raphael~Visoz,~\IEEEmembership{Senior Member,~IEEE,}
Didier~Le~Ruyet,~\IEEEmembership{Senior Member,~IEEE,}
Anvar~Tukmanov,~\IEEEmembership{Senior Member,~IEEE,}
Axel~M\"uller,~\IEEEmembership{Member,~IEEE,}
Alexander~Hamilton,~\IEEEmembership{Member,~IEEE,}
and~Matthew~Baker,~\IEEEmembership{Member,~IEEE}%
\thanks{
\setlength{\parindent}{0pt}

\textit{This paper has been submitted to IEEE Transactions 
on Vehicular Technology for possible publication.}\\

L.~Berrah and R.~Visoz are with Orange Research, Châtillon, 
France; L.~Berrah is also with CEDRIC, CNAM, Paris, France 
(e-mail: \texttt{lynda.berrah@orange.com}; 
\texttt{raphael.visoz@orange.com}).\\
D.~Le~Ruyet is with CEDRIC, CNAM, Paris, France 
(e-mail: \texttt{didier.leruyet@cnam.fr}).\\
A.~Tukmanov is with BT Group Plc, UK 
(e-mail: \texttt{anvar.tukmanov@bt.com}).\\
A.~M\"uller is with Nokia Networks, France 
(e-mail: \texttt{axel.mueller@nokia.com}).\\
A.~Hamilton and M.~Baker are with Nokia UK Limited 
(e-mail: \texttt{alexander.hamilton@nokia.com}; 
\texttt{matthew.baker@nokia.com}).}%
}



\maketitle


\begin{abstract}
Accurate channel modeling has become critical for evaluating multiple-input multiple-output (MIMO) performance, especially as 5G standardization matures and efforts toward 6G begin. Recent studies within the 3rd Generation Partnership Project (3GPP) have shown that the tapped delay line (TDL) model, currently used for performance testing, fails to capture the spatial propagation characteristics required for realistic MIMO evaluation. To address this limitation, the reduced clustered delay line (rCDL) model has been introduced as a more accurate alternative with manageable computational complexity, thereby enabling practical implementation in test equipment. This work investigates the rCDL through a comparative analysis with the legacy TDL. First, the angular characteristics of both models are examined. Then, their spatial profiles are compared with real-world measurements from a typical commercial deployment. The results reveal clear deficiencies in the TDL and show that the rCDL better matches measured propagation behavior. As a case study, channel state information (CSI) reporting performance is evaluated in single-user MIMO scenarios. The results show that, with appropriate simulation parameter settings, the rCDL enables clear discrimination between low- and high-resolution CSI reporting schemes, unlike the TDL. These findings confirm the relevance of the rCDL model for MIMO performance evaluation and support its use in current and future standardization efforts.
\end{abstract}

\begin{indexterms}
    Spatial channel modeling, rCDL, TDL, MIMO, CSI feedback, 5G/6G performance evaluation, 3GPP.
\end{indexterms}

\section{Introduction}

\IEEEPARstart{W}{ithin} the 3rd Generation Partnership Project (3GPP), the global standardization body for cellular systems, the responsibility for defining performance requirements for user equipment (UEs) and base stations (BSs) lies primarily with the Radio Access Network Working Group 4 (RAN4). RAN4 evaluates implementations under controlled conditions by specifying channel models, procedures, and test methodologies, thereby ensuring that devices meet the minimum performance targets prescribed in the standard. To this end, RAN4 uses the \textit{Tapped Delay Line (TDL)} channel model to define performance requirements due to its simplicity and low computational complexity.

In parallel, Radio Access Network Working Group 1 (RAN1), which is responsible for defining physical-layer techniques, conducts complex system-level evaluations that use the \textit{fast-fading Spatial Channel Model (SCM)} defined in TR 38.901 \cite{38901-Rel19}. RAN1 also resorts to link-level evaluations for specific topics, where the associated channel model is derived from the SCM and is known as the \textit{Clustered Delay Line (CDL)} model. Both RAN1 system-level and link-level models exhibit the “spatiality’’ property; {that is, they expose multiple, distinct angles of departure and arrival that remain stable as large-scale spatial characteristics.}

With the increasing sophistication of the multiple-input multiple-output (MIMO) techniques developed in RAN1, notably advanced high-resolution codebooks for channel state information (CSI) reporting \cite{3GPPCodebooks,38214-Rel19}, the RAN4 TDL channel model becomes unable to yield meaningful performance requirements. To address this limitation, 3GPP RAN4 recently initiated a Study Item (SI) \cite{3gppRP-241610} aimed at bridging the gap between RAN1 and RAN4 by adapting the RAN1 CDL channel models to the RAN4 use cases; that is, the resulting RAN4 channel model must preserve the “spatiality’’ property while maintaining low computational complexity and fast convergence to ease implementation in test equipment.

\subsection{Literature review}
MIMO channel models can be classified into \textit{analytical} and \textit{physical models} \cite{almers2007survey}. \textit{Analytical models} characterize the channel for each transmit-receive antenna link in a mathematical way without explicitly accounting for wave propagation. Moreover, the channel is defined at the antenna port level i.e., the antenna radiation pattern is directly included into the channel and is assumed omnidirectional. As an example, the 3GPP \textit{TDL} model specified in TS 38.101-4 \cite{38101-4-Rel19} defines a statistical power-delay profile (PDP) or a list of taps with delays and associated powers. Then, each tap coefficient is generated from a Rayleigh-Jakes distribution \cite{ImprovedJakes}. Extensions that incorporate antenna correlation based on the Kronecker model provided a rudimentary means to emulate MIMO behavior, but still lack spatial meaning \cite{METRAPaper}.

In contrast, \textit{physical models} explicitly characterize the interaction of the antenna pattern with the multipath propagation between the transmit and receive arrays, enabling a more accurate reproduction of radio propagation at the cost of higher computational complexity. They include \textit{ray-tracing methods}~\cite{RayTracing}, which rely on detailed environmental descriptions and extensive simulations, as well as \textit{geometry-based stochastic models}. The latter class retains the geometric interpretation of ray-tracing but replaces explicit scatterer positioning with stochastic descriptions based on probability distributions. The \textit{COST 273} model \cite{COST273ref} is the first model developed within this category that targets MIMO systems. It is based on the empirical observation that scatterers tend to form clusters with similar delays and angular characteristics, thereby introducing the concept of scatterer clusters. Later, \textit{COST 2100} \cite{COST2100ref}, generalized this framework to support large-scale MIMO, multi-user scenarios, and time-varying environments. {Recent works such as [1] and [2] introduce advanced cluster-based massive MIMO channel models that capture dual- and high-mobility effects by defining dynamic and static clusters.} Building on COST actions, 3GPP standardized the \textit{SCM} \cite{SCMref} for 3G systems, followed by the \textit{wideband extended SCM (SCME)} \cite{SCMEref} for beyond-3G. These models define large-scale parameters (e.g., delay and angular spreads) and small-scale parameters (e.g., per-cluster power, delay, directions of departure and arrival (DoDs/DOAs)) using tabulated probability distributions derived from measurement campaigns \cite{SCMPaper}. Subsequently, the \textit{WINNER} and \textit{WINNER II} models \cite{WINNERIref,WINNERIIref} were introduced, covering a broader frequency range, more deployment scenarios, and offering an improved elevation domain modeling. Alongside WINNER II, a reduced-complexity alternative, namely the \textit{CDL} model, was developed for link-level simulations to facilitate performance comparison and model calibration \cite{WINNERIIref}. The CDL fixes both large- and small-scale parameters to the expected values of the generic model. For 5G and beyond, 3GPP TR 38.901 further extends the WINNER II framework by defining a \textit{fast-fading SCM} together with its corresponding \textit{CDL} counterpart~\cite{38901-Rel19}. Recently, the aforementioned RAN4 SI on SCM for performance requirements has yielded the \textit{reduced CDL (rCDL)}, defined in TR 38.753 \cite{38753-Rel19}. A corresponding Work Item (WI) has now been approved \cite{3gppRP-252956}, marking the potential for standardization.

\subsection{Contributions}
Throughout the paper, we highlight the limitations of the TDL model, {which is either spatially agnostic or provides only limited monodirectional and tap-invariant spatial characterization}, and demonstrate that the rCDL model, when properly parameterized, represents a promising and accurate alternative for MIMO performance evaluation. {To the best of our knowledge, this channel model has not been previously analyzed in the open literature. While the rCDL was standardized in TR 38.753, the specification itself does not detail the design rationale, implications on channel statistical properties, performance evaluation, or practical implementation, as such information is typically withheld for the sake of competitivity. Our main contributions are summarized below.}
\begin{enumerate}
    \item We present a comprehensive overview of the 3GPP TDL
and CDL models employed for link-level simulations,
as standardized in TR 38.901, building on~\cite{38901-Rel19, SCMref}. {We derive the rCDL from the legacy RAN1 CDL formulation, detailing the rationale behind each design choice and its impact on the channel properties. Practical implementation guidelines are also provided.}
    \item We present a Bartlett-based DoA analysis of the TDL and CDL models, demonstrating that the CDL provides an accurate basis for MIMO performance evaluation, whereas the TDL model does not.
    \item We conducted channel measurements in a commercial MIMO deployment from which {we extracted post-equalization SINR profile per spatial layer}. The results show that the TDL model fails to capture real-world propagation characteristics, while the CDL model matches the measured behavior.
    \item We provide a detailed overview of the 3GPP Type-I (low-resolution) and enhanced Type-II (high-resolution) codebooks for CSI reporting, with emphasis on their underlying design principles. A practical precoder selection strategy based on these codebooks is also described, along with a physical-layer abstraction framework for link-level performance evaluation.
    \item Using the proposed framework, we present spectral efficiency results {under TDL, CDL, and rCDL models, showing that the rCDL enables clear discrimination between low- and high-resolution codebooks when properly parameterized. Critical parameters governing the degree of discrimination are further discussed.}
\end{enumerate}

\subsection{Paper Organization}
The remainder of the paper is organized as follows.
Sec. II reviews the TDL model and examines its spatial properties.
Sec. III provides background on the CDL model, analyzes its spatial profile, and derives the rCDL variant.
Sec. IV presents field measurements conducted in a commercial MIMO deployment.
Sec. V discusses CSI reporting based on 3GPP codebooks as an application of MIMO performance evaluation under TDL and rCDL models.
Sec. VI concludes the paper.

\section{Tapped Delay Line (TDL) Model}
\label{S.TDL}
\subsection{Description of the TDL Channel Model}
Consider a single-input single-output (SISO) non-line-of-sight (NLOS) channel, the passband transmit signal is
 \begin{equation} 
 e(t) = \Re{ \bigl\{ x(t) \ e^{j2\pi f_c t} \bigr\}}, 
 \end{equation}
where $x(t)$ is the baseband transmit signal, $f_c$ is the carrier frequency, and $\Re\{\cdot\}$ denotes the real part. Assuming a finite bandwidth, i.e., discrete propagation delays, the passband received signal can be modeled as a superposition of multipath components (MPCs), that is
\begin{equation} 
r_{pb}(t) = \sum_{n=1}^{N} \sum_{m=1}^{M} \alpha_{n,m} \ \Re\Bigl\{ e^{j \phi_{n,m}} x\bigl(t-\tau_{n,m}(t)\bigr) e^{j2\pi f_c \bigl( t - \tau_{n,m}(t)\bigr)}\Bigr\}, \label{e.SignalPB} 
\end{equation}
with MPC delays $\tau_{n,m}(t) \in [\tau_n(t)-\Delta \tau, \tau_n(t)+\Delta \tau]$, where $\Delta \tau$ is small enough so that $x(t-\tau_{n,m}(t)) \approx x(t-\tau_{n}(t))$. We also assume that the delay variation does not significantly distort the transmit signal, i.e., $x(t-\tau_n (t)) \approx x(t-\tau_n)$, {which is true when $(\tau_n(t)-\tau_n)\ll1/B$, where $B$ is~the~transmit~signal bandwidth}. Under these assumptions, the baseband received signal is
\begin{equation} 
r_{bb}(t) = \sum_{n=1}^{N} \sum_{m=1}^{M} \alpha_{n,m} e^{j \phi_{n,m}} e^{- j2\pi f_c \tau_{n,m}(t)} x(t-\tau_{n}). \label{e.SignalBB} \end{equation} 
Equivalently, \eqref{e.SignalBB} can be written as a linear time-varying convolution as follows
\begin{equation} 
r_{bb}(t) = h(t,u) \ast x(u) \Big|_{u=t}, 
\end{equation}
with the channel impulse response (CIR) given by 
\begin{equation} h(t,\tau) = \sum_{n=1}^{N} c_n(t) \delta(\tau-\tau_{n}), \label{e.CIR} \end{equation}
where $\delta(\cdot)$ is the Dirac delta function and the complex gain associated with tap $n$ is written as
\begin{equation}
c_n(t) = \sum_{m=1}^{M} \alpha_{n,m} e^{j \phi_{n,m}} e^{- j2\pi f_c \tau_{n,m}(t)}. \label{e.MPC} 
\end{equation} 
The channel is described by $N$ taps, each being the superposition of $M$ MPCs arriving with nearly the same delays. $\alpha_{n,m}$ and $\phi_{n,m}$ denote the amplitude and initial phase of the $m$-th ray belonging to tap $n$, respectively. At a given time $t$, for large $M$, by the central limit theorem $c_n(t)$ is well modeled by a complex Gaussian process; consequently, $|c_n(t)|$ is Rayleigh distributed. Empirical measurements in NLOS scenarios support the Rayleigh fading assumption. In the standard, the TDL model in \eqref{e.CIR} is specified by the tables in \cite[Appendix~B]{38101-4-Rel19}, each corresponding to a frequency band and profile. These tables define the number of taps and their associated delays and average powers. Furthermore, the standard specifies that the tap coefficients exhibit Rayleigh fading for NLOS components.

Consider the receiver moving with velocity vector $\mathbf{v}$, and let $\mathbf{u}_{n,m}$ be the unit vector corresponding to the inverse direction of arrival of the ray $m$ in cluster $n$, the delay evolves as
\begin{equation}
    \tau_{n,m}(t) = \tau_{n,m} - \frac{\mathbf{u}_{n,m}.\mathbf{v}}{c}t.
\end{equation}
Let $\varphi_{n,m} \coloneqq (\phi_{n,m} - 2\pi f_c\tau_{n,m}(t))$ and the maximum Doppler shift $f_D \coloneqq v/\lambda_0$, where $v$ is the speed and $\lambda_0$ denotes the wavelength. Assuming two-dimensional scattering, i.e., the angular domains reduce to the azimuthal domain, and let $\theta_{n,m}$ be the angle between $\mathbf{u}_{n,m}$ and $\mathbf{v}$, the gain for tap $n$ in \eqref{e.MPC} can be re-written as 
\begin{equation}
    c_n(t) =  \sum_{m=1}^{M} \alpha_{n,m} e^{j \varphi_{n,m}}  e^{j2\pi f_D t \cos(\theta_{n,m})}.
\end{equation}
This channel time variation is characterized by its power spectral density (PSD), also called the Doppler spectrum. With a large number of MPCs and i.i.d. angles $\theta_{n,m}\sim\mathcal{U}[0,2\pi)$, the normalized temporal autocorrelation is given as \cite{ClarkePaper}    
\begin{equation}
    R(\tau) =   J_0 (2\pi f_D \tau),
    \label{e.autocorrelation}
\end{equation}
where $J_0(.)$ is the zero-order Bessel function of the first kind. 
By taking the Fourier transform of \eqref{e.autocorrelation}, the Doppler spectrum reduces to the so-called Jakes spectrum
\begin{equation}
    S(f) = \frac{1}{\pi f_D \sqrt{1-\left(f/f_D\right)^2}}, \quad |f|<f_D.
    \label{e.psdDoppler}
\end{equation}
and $S(f) = 0$ for $|f|\ge f_D$. This is also referred to as the Clarke's model. Channel coefficients can be generated by means of a sum-of-sinusoids (SoS) model as defined below.

\noindent \textit{Definition: Clarke's SoS implementation~\cite{ClarkePaper}}
    \begin{equation}
        c_n(t) = \frac{1}{\sqrt{M_0}} \sum_{m=1}^{M_0} e^{j\left( 2\pi f_D t \cos(\theta_{n,m})+ \varphi_{n,m}\right)}
        \label{e.Clarke}
    \end{equation}
where $M_0$ denotes the number of Doppler-shifted sinusoids. Here, $\varphi_{n,m}$ and $\theta_{n,m}$ are statistically independent and uniformly distributed on $[0,2\pi)$. The value of $M_0$ must be large enough to approximate Rayleigh statistics.  {Based on \eqref{e.Clarke}, Jakes derived a simplified simulator in~\cite{JakesBook}, followed by several alternative implementations; a commonly used one is described in~\cite{ImprovedJakes}.}

\subsection{Extension to MIMO Scenarios}
The TDL model in \eqref{e.CIR} extends to MIMO systems by treating each transmit–receive antenna pair as an independent SISO channel, this by replacing scalar taps with matrices as follows
\begin{equation}
    \textbf{H}(t,\tau) = \sum_{n=1}^N \widetilde{\mathbf{C}}_n(t) \ \delta(\tau-\tau_n),
\end{equation}
where $\widetilde{\mathbf{C}}_n(t) \in \mathbb{C}^{N_R \times N_T}$, with $N_R$ and $N_T$ denoting the numbers of receive and transmit antennas, respectively. 

Spatial correlation can be imposed on the entries of $\widetilde{\mathbf{C}}_n(t)$ to match the desired channel properties. Let $\widetilde{\mathbf{c}}_n(t) \coloneqq \operatorname{vec}\big(\widetilde{\mathbf{C}}_n(t))\big)$, where $\operatorname{vec}(\cdot)$ stacks the $N_R \times N_T$ matrix into a column vector. The coefficients for tap $n$ are generated as
\begin{equation}
    \mathbf{\widetilde{c}}_n(t) = \left(\mathbf{R}_n \right)^{1/2} \mathbf{c}_n(t),
\end{equation}
where $\left(\mathbf{R}_n \right)^{1/2} \in \mathbb{C}^{N_R N_T \times N_R N_T}$ is any square-root decomposition of the spatial correlation matrix $\mathbf{R}_n$, and $\mathbf{c}_n(t) \in \mathbb{C}^{N_R N_T \times 1}$ has i.i.d. Rayleigh fading coefficients generated via the SoS method at time $t$. {Predecessors of the 3GPP models define the correlation matrices as the Kronecker product of the transmit- and receive-side correlation matrices, derived independently from the power azimuth spectrum (PAS) under a uniform linear array (ULA) assumption. For instance, a rectangular PAS yields a Bessel-function relationship between the correlation coefficient and the array spacing~\cite{METRAPaper}. Building on these simplifications, 3GPP defines the correlation matrices in~\cite{38101-4-Rel19} under the following additional assumptions.}
\begin{itemize}
        \item {\textit{Identical correlation matrices for all taps:}} $2\times2$ correlation matrices are adopted, assumed identical across all taps, for settings with $2$ transmit and receive antennas. In realistic propagation environments, this assumption is generally not valid, as different scatterers typically exhibit distinct spatial profiles~\cite{3gppR4-071317}.
        \item {\textit{Real values:}} Phase is disregarded and only the magnitudes are considered. This potentially obscures important spatial information, since multiple distinct power spectra can yield identical magnitude values~\cite{3gppR4-072223}. 
        \item {\textit{Exponential interpolation:}} The model is extended to $4\times4$ correlation matrices via an ad-hoc exponential decay interpolation~\cite{3gppR4-080127}, and subsequently to $8\times8$~\cite{3gppR4-1708533}. However, the validity of this dimensional extension with respect to the original measurements remains unclear~\cite{Durgin}. 
\end{itemize}

\subsection{Spatial Profile}
\label{S.TDLSpatialProfile}
We define the receive spatial profile of a channel realization as the receive signal power obtained when discrete Fourier transform (DFT) beams are applied over varying DoAs. This simple approach, referred to as Bartlett analysis, is used to estimate the power azimuth spectrum of the channel model. More complex high-resolution methods, such as MUSIC and ESPRIT, also exist; however, Bartlett approach is sufficient to provide insights into the spatial characteristics of the TDL model.  {According to~\cite{van2002optimum}, the Bartlett scan output is given as
\vspace{-0.5em}
\begin{equation}
    P_{\mbox{Bartlett}}(\theta) = \textbf{a}^H(\theta) \textbf{R}_{rx} \, \textbf{a}(\theta),
    \label{e.BartlettOutput}
\end{equation}
\noindent where $\textbf{a}(\theta)$ denotes the array steering vector corresponding to direction $\theta$, and $\textbf{R}_{rx}=\frac{1}{\beta}\mathbf{H}\mathbf{H}^\mathrm{H}$ represents the normalized spatial correlation matrix at the receiver side. Here, $\mathbf{H}$ denotes the channel realization at the resource element (RE) level, and $\beta=tr(\mathbf{H}\mathbf{H}^\mathrm{H})$ ensures unit total power, thereby enabling fair comparison across different channel realizations.} 

\begin{figure}[b!]
    \centering
    \includegraphics[width=0.63\linewidth, trim=10 0 30 0, clip]{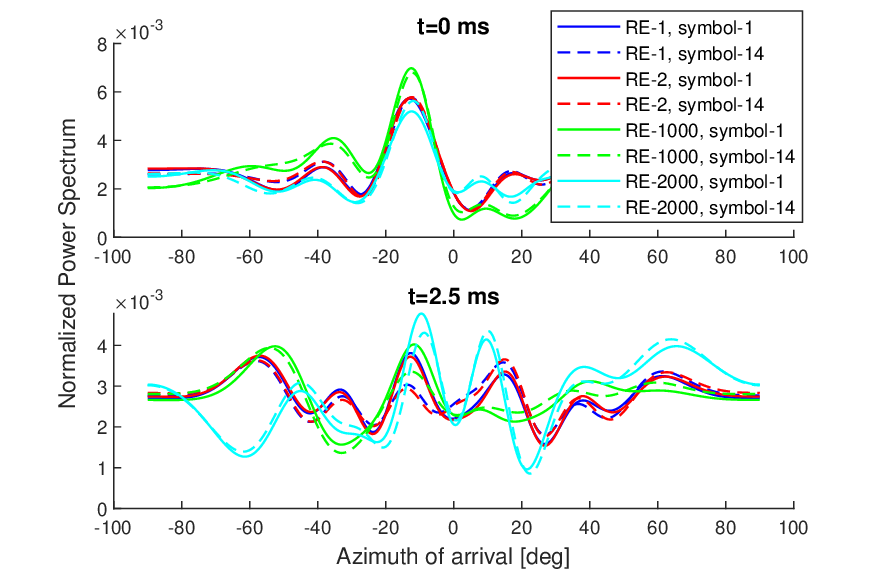}
    \caption{Bartlett DoA analysis comparing RE-wise temporal evolution of the TDLC300-100 model with low antenna correlation; {normalization is applied with respect to the total power.} High spatial randomness and rapid per-RE angular decorrelation are observed.}
    \label{f.R4-2411300Fig2_BartlettDoATDLC}
\end{figure}

In Fig.~\ref{f.R4-2411300Fig2_BartlettDoATDLC}, we examine the spatial profile of TDL models without antenna correlation (i.e., low correlation in 3GPP parlance).
We consider a MIMO system with $8$ transmit and receive antenna ports at a carrier frequency $f_c=3.5$~GHz, with a system bandwidth of $40$~MHz, and a sub-carrier spacing of $30$~kHz. We adopt a TDL-C profile with a delay spread of $300$~ns and a maximum Doppler frequency of $100$~Hz, corresponding to a UE speed of $30$~Km/h. This configuration is referred to as TDLC300-100 and is defined in~\cite[Table B.2.1.1-4]{38101-4-Rel19}. 

For TDLC300-100 with low correlation, we observe that the large-scale spatial preference of the per-RE channels becomes mostly decorrelated after $2.5$~ms. Although the speed of change of the spatial profile can be theoretically controlled using the maximum Doppler frequency, it is here misaligned with expectations. Moreover, the large-scale spatial properties are hardly repeatable between implementations, and expected mid-term stable directions are not present. Following the aggregation of uncorrelated SISO channel coefficients into a MIMO channel model, we also note that realizations of the TDL low model are almost surely full rank. This excessive randomness of the spatial profile (illustrated in Fig.~\ref{f.R4-2411300Fig2_BartlettDoATDLC}) poses challenges for precoding and CSI reporting procedures, which were recognized in 3GPP. 

\begin{figure}[h]
    \centering
    \includegraphics[width=0.63\linewidth, trim=10 0 30 0, clip]{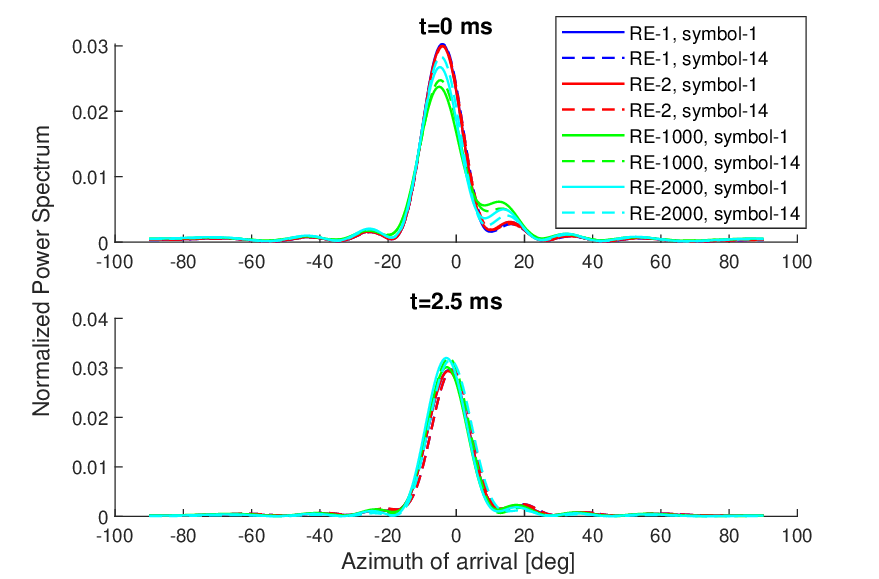}
    \caption{Bartlett DoA analysis comparing RE-wise temporal evolution of the TDLC300-100 model with medium-A antenna correlation; {normalization is applied with respect to the total power.} A single DoA, constant across time and REs, is observed.}
    \label{f.R4-2411300Fig6_BartlettDoATDLC}
\end{figure}

{To address this issue, antenna correlations were introduced based on the Kronecker model for CSI performance requirements definition. 3GPP defines a real-valued ULA covariance matrix assumed identical across all delay taps (cf. Sec.~II-B), with spatial correlation strongest at small antenna spacings. As a result, the Bartlett scan output in \eqref{e.BartlettOutput} is maximized when all antennas add fully in phase, a condition occurring at broadside. Moreover, since the same spatial correlation matrix is applied to every delay tap, the resulting spatial spectrum exhibits a single dominant peak. This {monodirectional tap-invariant behavior} is illustrated in Fig.~\ref{f.R4-2411300Fig6_BartlettDoATDLC} for TDLC300-100 with medium correlation \cite[Sec.~B.2.3.1.2]{38101-4-Rel19}, where a consistently exaggerated spatial directivity toward broadside is observed across all realizations, REs, and delay taps. Consequently, TDL realizations employing such antenna correlation are highly rank deficient;} increasing the antenna number above $3$ does not improve spatial diversity due to exceedingly small inter-antenna correlation coefficients. Additionally, both signal and interference are consistently received from the broadside direction, independently of the chosen precoder. 

\section{Clustered Delay Line (CDL) Model}
\label{S.CDL}
\begin{figure}[htb]
    \centering
    \includegraphics[width=0.6\linewidth, trim = 3.8cm 1.6cm 3.2cm 1.6cm, clip]{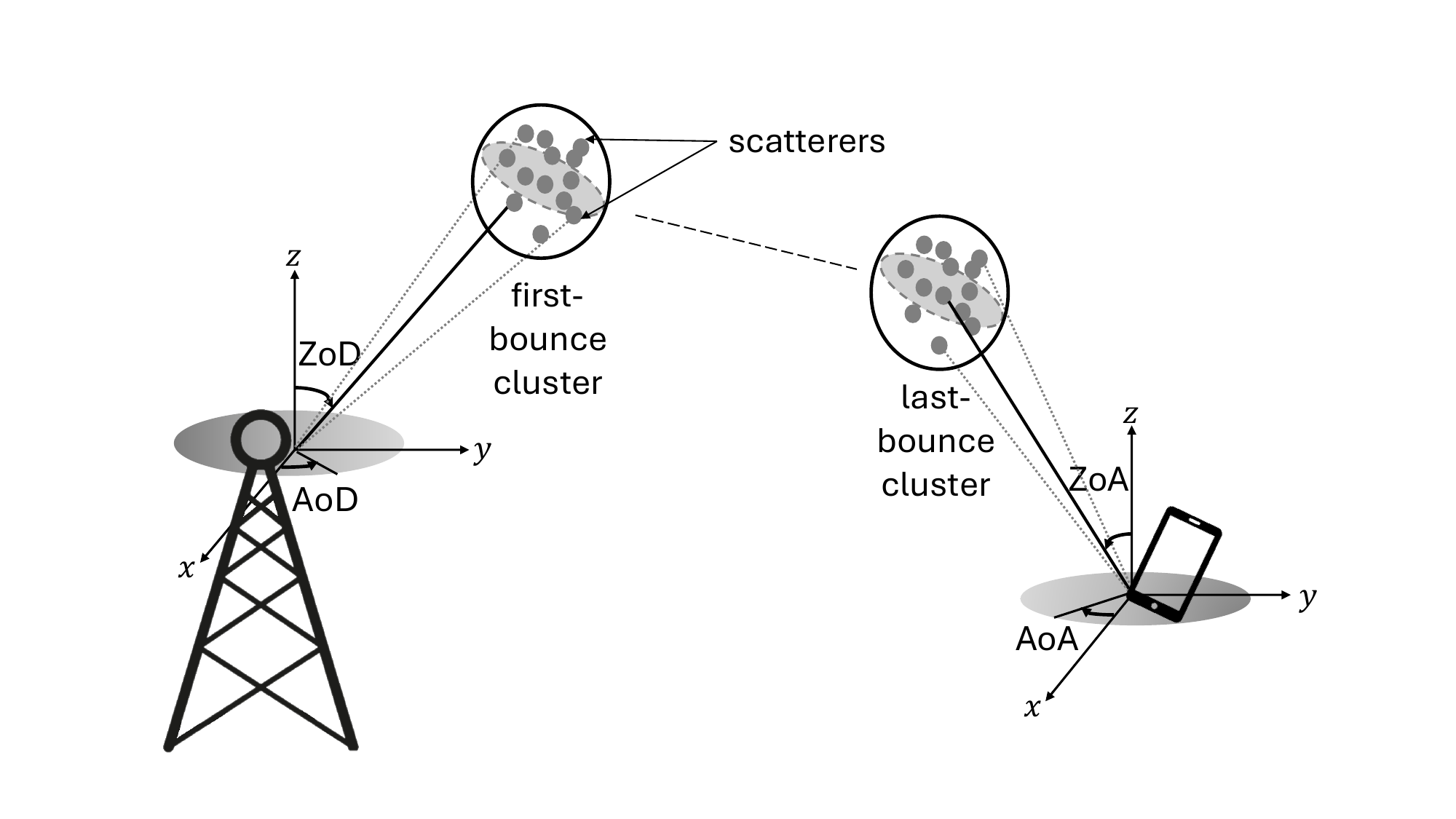}
    \caption{Illustration of the CDL model. The cluster is depicted as an ellipsoid that groups scatterers, and has a mean azimuth/zenith of departure/arrival, denoted AoD, ZoD, AoA, and ZoA, respectively.}
    \label{f.CDLIllustration}
\end{figure}

 The CDL model represents the propagation channel as a superposition of multiple distinct clusters, each corresponding to a dominant scattering region characterized by a specific propagation delay, average power, and mean angles of departure and arrival. Within each cluster, several rays or MPCs share the same delay but exhibit small variations in their angles, which are distributed around the cluster mean according to predefined angular spreads \cite{SCMPaper}. Fig. \ref{f.CDLIllustration} illustrates a cluster composed of multiple scatterers, modeled as first- and last-bounce ellipsoids, as observed from the BS and UE.

The CDL model is defined by fixed large- and small-scale parameters (LSPs and SSPs, respectively). The LSPs include the delay spread and angular spreads, while the SSPs comprise cluster power, delay, DoAs/DoDs, cross-polarization power ratio (XPR), and the per-cluster angular spreads in both azimuth and zenith, as illustrated in Fig.~\ref{f.CDLIllustration}. These per-cluster parameters are set to median values derived from the generic stochastic fast-fading SCM model, which defines statistical distributions for LSPs and SSPs \cite[Sec. 7.5]{38901-Rel19}, based on measurement campaigns in various environments (e.g., Urban Macro (UMa), Indoor Factory (InF)). The CDL model is fully specified through tables in \cite[Sec. 7.7]{38901-Rel19} for different profiles: CDL-A/B/C for NLOS and CDL-D/E for LOS scenarios. 
\vspace{-0.15 cm}
\subsection{Description of the CDL Channel Model}
The model considers $N$ clusters, each comprising $M$ rays. The channel coefficient in NLOS scenario for transmit antenna~$s$, receive antenna $u$, and cluster $n$ at time $t$ is
    \begin{multline}
 H_{u,s,n}(t) = \sqrt{\frac{P_n}{M}}\sum_{m=1}^M \mathbf{F}_{rx,u}(\theta_{n,m,ZoA},\phi_{n,m,AoA})^T  
  \mathbf{M}_{pol} \ \mathbf{F}_{tx,s}(\theta_{n,m,ZoD},\phi_{n,m,AoD}) \\ \exp \left(\frac{j2\pi}{\lambda_0} \mathbf{\hat{r}}_{rx,n,m}^T. \mathbf{\bar{d}}_{rx,u}\right)  \exp \left(\frac{j2\pi}{\lambda_0} \mathbf{\hat{r}}_{tx,n,m}^T. \mathbf{\bar{d}}_{tx,s}\right) \exp \left(\frac{j2\pi}{\lambda_0} \mathbf{\hat{r}}_{rx,n,m}^T. \mathbf{\bar{v}} \,t\right).
    \label{e.CDLModel}
\end{multline}
Eq. \eqref{e.CDLModel} is expressed in spherical coordinates, where $\theta$ and $\phi$ denote the zenith (ZoD, ZoA) and azimuth (AoD, AoA) angles of departure and arrival, respectively (with $\theta = 90^\circ$ pointing toward the horizon). $P_n$ represents the power of the $n$-th cluster. For clarity, we omit indices in the following description. The vectors $\mathbf{F}_{tx}$ and $\mathbf{F}_{rx}$ correspond to the transmit and receive antenna field patterns, respectively, while $\mathbf{M}_{pol}$ 
denotes the polarization coupling matrix. The unit vectors $\mathbf{\hat{r}}_{tx}$ and $\mathbf{\hat{r}}_{rx}$ define the directions of propagation from the transmitter and toward the receiver, and the vectors $\mathbf{\bar{d}}_{tx}$ and $\mathbf{\bar{d}}_{rx}$ specify the positions of the transmit and receive antenna elements relative to the center of the panels. Finally, $\mathbf{\bar{v}}$ represents the UE velocity vector. In the following, we provide a detailed description of each term in \eqref{e.CDLModel}.\\

\noindent\subsubsection{Angle Definition:} The AoD of ray $m$ in cluster $n$ is
\begin{equation}
    \phi_{n,m,AoD} = \phi_{n,AoD} + c_{ASD} \alpha_m,
\end{equation}
where $\phi_{n,{AoD}}$ is the cluster mean AoD and $c_{{ASD}}$ is the per-cluster azimuth spread of departure (ASD), both tabulated in \cite{38901-Rel19}, e.g., Table 7.7.1-3 for CDL-C. The ray offset angles $\alpha_m$ are also fixed by 3GPP \cite[Table 7.5-3]{38901-Rel19}. Other angles (i.e., AoA, ZoA, and ZoD) are defined analogously using their respective cluster means and per-cluster spreads. Within each cluster, the previously defined angles are then randomly coupled to form the final ray directions.\\

\noindent\subsubsection{Antenna Modeling:}
\label{s.AntennaModeling}The field pattern describes the spatial distribution of the radiated electromagnetic field from a given antenna element. In polar coordinates, it is given as
\begin{equation}
    \mathbf{F}''(\theta'',\phi'') = \begin{bmatrix}
        F_{\theta''}''(\theta'',\phi'') \\
        F_{\phi''}''(\theta'',\phi'')
    \end{bmatrix},
    \label{e.FieldsDP}
\end{equation}
where $F^{\prime\prime}_{\theta^{\prime\prime}}$ and $F^{\prime\prime}_{\phi^{\prime\prime}}$ denote the co-polar and cross-polar components of the field in the directions of the spherical unit vectors $\hat\theta$ and $\hat\phi$, respectively. The corresponding power radiation pattern is given as
\begin{equation}
    A''(\theta,\phi) =|F''_{\theta''}(\theta'',\phi'')|^2+|F''_{\phi''}(\theta'',\phi'')|^2.
\end{equation}
For a single purely vertically polarized antenna located at the origin of the coordinate system, the field pattern simplifies to
\begin{equation}
    \mathbf{F''}(\theta'',\phi'') =  \begin{bmatrix}
        \sqrt{A''(\theta'',\phi'')} \\
       0
    \end{bmatrix},
    \label{e.FieldLCSpp}
\end{equation}
with $A''(\theta'',\phi'')=0$~dBi  for an isotropic radiator, and defined according to formula in~\cite[Table 7.3-1]{38901-Rel19} for a sector antenna. 

To generalize the definition to arbitrary antenna orientations, we introduce a hierarchy of coordinate systems. A global coordinate system (GCS) provides a common reference for the transmitter and receiver, in which all the terms in \eqref{e.CDLModel} are expressed in the standard. The local coordinate systems (LCS$'$) are obtained by rotating the GCS according to the orientation of each antenna panel, described by the angles $(\alpha, \beta, \gamma)$. Here, $\alpha$ is the bearing angle (rotation about the $z$-axis), $\beta$ is the downtilt angle (rotation about the $y$-axis), and $\gamma$ is the slant angle (rotation about the $x$-axis). Separate LCS$'$ are defined for the BS and UE panels.  To account for polarization slant, we introduce double-primed local coordinate systems (LCS$''$). Each LCS$''$ is obtained by rotating the corresponding LCS$'$ about the $x$-axis by the polarization slant angle $\zeta$, assuming the antenna panel lies in the $(y',z')$ plane. Hence, for each polarization of each panel (BS and UE), we define a separate LCS$''$, resulting in four double-primed coordinate systems. Within each LCS$''$, the antenna elements of the considered polarization are treated as vertically polarized, so that the formula in \eqref{e.FieldsDP} applies.  The GCS-to-LCS transformations are detailed in the Appendix.\\

\noindent\subsubsection{Polarization Coupling:}
The polarization matrix $\mathbf{M}_{pol}$ describes how the polarization changes on the propagation path from the transmitter to the receiver. It is modeled based on random coefficients as
\begin{equation}
    \mathbf{M}_{pol} = \begin{bmatrix}
        \exp{(j \Phi_{n,m}^{\theta \theta})} &  \frac{1}{\sqrt{\kappa_{n,m}}}\exp{(j \Phi_{n,m}^{\theta \phi})} \\
         \frac{1}{\sqrt{\kappa_{n,m}}} \exp{(j \Phi_{n,m}^{\phi \theta})} &  \exp{(j \Phi_{n,m}^{\phi \phi})}
    \end{bmatrix}
    \label{e.PolarMatrix}
\end{equation}
Here, the phases for the four polarization combinations $\left\{\Phi_{n,m}^{\theta \theta}, \Phi_{n,m}^{\phi \theta}, \Phi_{n,m}^{\theta \phi}, \Phi_{n,m}^{\phi \phi}\right\}$ are drawn from a uniform distribution on $[-\pi,\pi)$. The coefficient $\kappa_{n,m}$ is the XPR for ray $m$ in cluster $n$, it quantifies the separation between the two polarized channels due to different orientations. CDL models use a fixed XPR for all clusters and rays, tabulated in~\cite[Table 7.7.1-3]{38901-Rel19} for CDL-C.\\

\noindent\subsubsection{Phase Shifts across Antenna Arrays:}
\label{s.PhaseShifts}
\begin{figure}[t!]
    \centering
    \includegraphics[width=0.6\linewidth, trim = 0.38cm 0.6cm 1.4cm 0.17cm, clip]{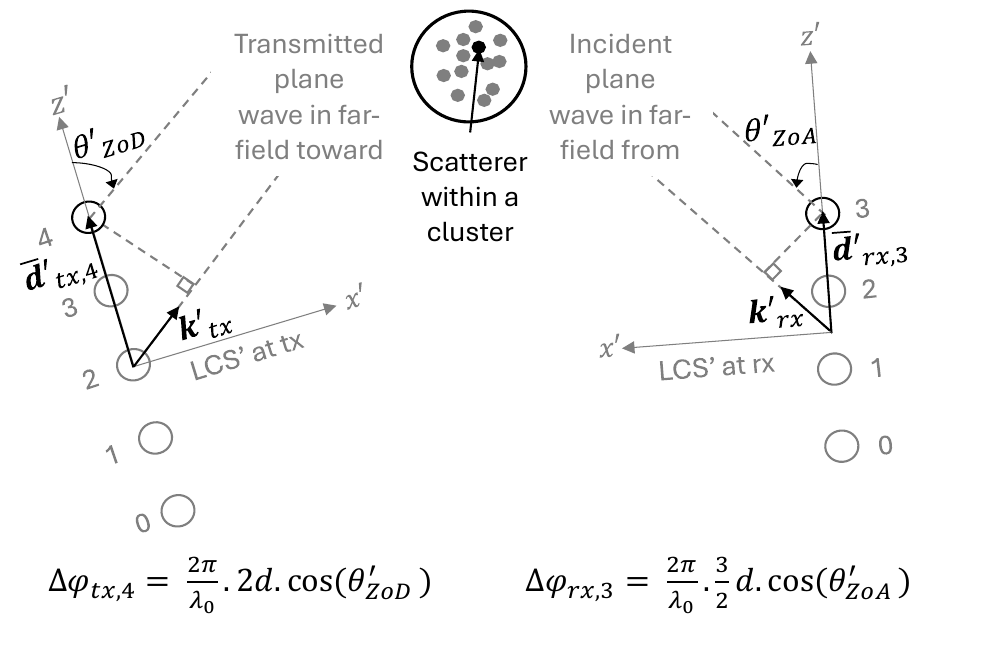}
    \caption{Phase shifts for a given ray within a scatterer cluster for a given transmit and receive uniform linear arrays (ULA), illustrating the effect of antenna element positions on the received signal phase. \vspace{-0.2 cm}}
    \label{f.PhaseShifts}
\end{figure}

Here, we express the phase shift associated with the  scatterer $m$ in cluster~$n$ when considering transmit antenna element $s$ and receive antenna element~$u$, relative to a hypothetical ray assuming that the antenna elements are located at the centers of the transmit and receive panels. Assuming that the distance between the scatterer and the array is much larger than the array aperture, the wave arriving at or departing from any scatterer can be approximated as a plane wave. Under this far-field approximation, the phase shift for the transmit antenna–scatterer link~is
\begin{equation}
\Delta \varphi_{tx,s,n,m} =\exp \left(\frac{j2\pi}{\lambda_0} \mathbf{\hat{r}}_{tx,n,m}^T. \mathbf{\bar{d}}_{tx,s}\right),
\end{equation}
and for the scatterer–receive antenna link as
\begin{equation}
\Delta \varphi_{rx,u,n,m} = \exp \left(\frac{j2\pi}{\lambda_0} \mathbf{\hat{r}}_{rx,n,m}^T. \mathbf{\bar{d}}_{rx,u}\right).
\end{equation}
The wave vector for the link between transmit antenna $s$ and scatterer $m$ within cluster $n$ is expressed as
\begin{equation}
  \mathbf{{k}}_{tx,n,m} =   \frac{2\pi}{\lambda_0} \mathbf{\hat{r}}_{tx,n,m} = \frac{2\pi}{\lambda_0} \begin{bmatrix}
\sin \theta_{n,m,ZoD} \cos \phi_{n,m,AoD} \\
\sin \theta_{n,m,ZoD} \sin \phi_{n,m,AoD} \\
\cos \theta_{n,m,ZoD}
\end{bmatrix},
\label{e.WaveVector}
\end{equation}
and the location vector of the transmit antenna $s$, expressed in the BS LCS$'$ as
\begin{equation}
    \mathbf{\bar{d}'}_{tx,s} = \begin{bmatrix}
        0 & n_s d_y - y_c&
        m_s d_z - z_c
    \end{bmatrix}^T,
    \label{e.locationVector}
\end{equation}
where $(d_y,d_z)$ denote the inter-element spacings along the horizontal and vertical dimensions, respectively, and $(n_s,m_s)$ denote the antenna element indices. The origin of the GCS and LCS$'$ is assumed to be located at the center of the antenna panel. Accordingly, in \eqref{e.locationVector}, $(y_c,z_c)\coloneqq \bigl(\tfrac{N_y-1}{2}d_y,\tfrac{N_z-1}{2}d_z\bigr)$ represent the coordinates of the panel center, where $N_y$ and $N_z$ denote the number of antenna elements along the $y$- and $z$-axes, respectively. However, the absolute choice of the GCS origin for the CDL model is inconsequential {(up to an additive constant phase term, which has no impact)}, since the phase term depends only on the relative orientation between the wave propagation direction and the antenna element displacement vector, provided both are expressed in the same coordinate system. Consequently, the scalar product between \eqref{e.WaveVector} and \eqref{e.locationVector} can be evaluated directly in the LCS$'$. To this end, the wave propagation vector is transformed to LCS$'$ according to
\begin{equation}
    \mathbf{\hat{r}}'_{tx,n,m} = \mathbf{R}^{T} \ \mathbf{\hat{r}}_{tx,n,m},
\end{equation}
where the rotation matrix $\mathbf{R}$ is detailed in the Appendix. The phase shifts at the receiver are computed analogously by considering the arrival angles in \eqref{e.WaveVector} and the antenna locations on the receive panel in \eqref{e.locationVector}. Fig. \ref{f.PhaseShifts} illustrates the phase shifts in a two-dimensional propagation scenario.\\

\noindent\subsubsection{Doppler Effect:}
\label{s.DopplerTerm}The temporal evolution of the CIR is modeled based on a Doppler shift that depends on the velocity $v$, the user direction of motion $(\phi_v,\theta_v)$, and the arrival angles as in Sec. \ref{S.TDL}. It is given as 
\begin{equation}
\frac{\mathbf{\hat{r}}_{rx,n,m}^T. \mathbf{\bar{v}}}{\lambda_0}, \text{ with }    \mathbf{\bar{v}} = v \begin{bmatrix}
        \sin \theta_v \cos \phi_v \! \! & \!  \sin \theta_v \sin \phi_v \! \! & \!  \cos \theta_v
    \end{bmatrix}^T.
\end{equation}
\newpage
\noindent\subsubsection{Antenna Numbering:}
The channel coefficients calculated based on \eqref{e.CDLModel} are stacked from bottom to top, column by column, as described in \cite{388897-Rel18} starting with the co-polarized then the cross-polarized elements. This is important as the precoder structure as defined by 3GPP assumes this numbering.

\subsection{Spatial Profile}

\begin{figure}[b!]
    \centering
    \includegraphics[width= 0.63\linewidth, trim=10 0 30 8, clip]{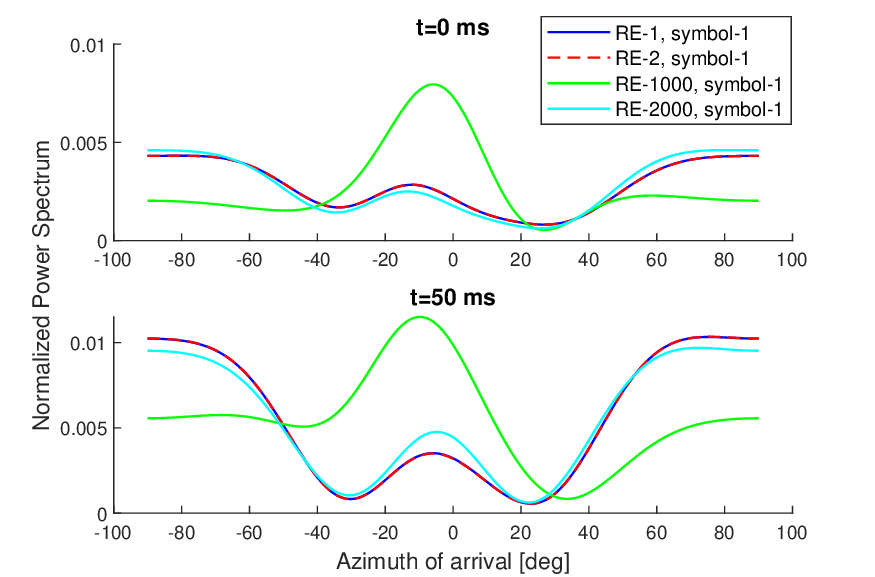}
    \caption{Bartlett DoA analysis comparing RE-wise temporal evolution of the CDL-C UMa model with $365$~ns delay spread and $100$~Hz Doppler frequency; {normalization is applied with respect to the total power.} Multiple mid-term stable DoAs are observed.}
    \label{f.R4-2411300Fig8_BartlettDoACDLC}
\end{figure}
\label{S.CDLSpatialProfile} A Bartlett-style spatial profile analysis, identical to that described in Sec.~\ref{S.TDLSpatialProfile}, is applied to the CDL model. We consider the CDL-C profile (UMa scenario) with a delay spread of $365$~ns and a maximum Doppler shift of $100$~Hz in order to obtain results that are comparable with those of the TDL model. The same carrier frequency and bandwidth are used for both models. Unlike the TDL model, which defines the channel directly at the antenna port level, here we consider $128$ transmit antenna elements mapped to $8$ antenna ports as in~\cite{38827-Rel16}, and $8$ receive antenna ports. 
The resulting spatial profile in Fig.~\ref{f.R4-2411300Fig8_BartlettDoACDLC} demonstrates that the CDL model captures the spatial characteristics of practical MIMO deployments, i.e., the presence of multiple mid-term stable DoAs. An additional advantage of the CDL model is that the channel rank naturally follows from the antenna modeling and cluster configuration. Consequently, the rank is inherent to the propagation and antenna assumptions, rather than being imposed through external antenna correlation models.


\subsection{Derivation of the Reduced CDL (rCDL) Model}
{
Based on the RAN1 CDL-C model described in Sec.~III-A, the rCDL model was derived for laboratory RAN4 MIMO conformance testing~\cite{38753-Rel19}. In such testing scenarios, real-time channel emulation hardware from multiple vendors is employed. Furthermore, since practical testing durations are typically limited to approximately 20 seconds per SNR point, different hardware implementations must exhibit rapid convergence to ensure comparable performance results. As a representative example, in the PMI reporting procedure, the UE is required to demonstrate a precoding gain relative to random precoding that exceeds the minimum threshold specified in~\cite[Sec.~8.3]{38101-4-Rel19}. Consequently, the design choices adopted for the rCDL model primarily aim to reduce inter-vendor convergence time. Accordingly, the following modifications are applied to the RAN1 CDL-C to derive the rCDL-C model.}

\vspace{0.15 cm}

\noindent \subsubsection{Randomness Reduction Framework:}
{By agreeing on certain CDL parameters that strongly govern the initial state and time evolution of the channel, the inter-implementation convergence time is expected to be significantly reduced. This randomness reduction framework encompasses the use of: }
\begin{itemize}
    \item {\textbf{Fixed ray coupling:} The transmit/receive ray coupling patterns, as well as the azimuth/zenith coupling, are fixed so that all hardware implementations start from the same spatial profile, thereby avoiding the need for many independent random draws to converge.} The couplings are provided in \cite[Table 5.1.3.2-1]{38753-Rel19}. 
    \item {\textbf{Fixed initial phases:} To further reduce inter-vendor convergence time, the initial phases of the sub-paths associated with cross-polarization effects are fixed to ensure consistent small-scale fading evolution across different hardware implementations.} The chosen values are provided in \cite[Table 5.1.3.2-2]{38753-Rel19}.
\end{itemize}

\vspace{0.1 cm}

\noindent \subsubsection{Angular Scaling Procedure:}
\label{s.AngleScaling}The angles of departure and arrival of the RAN1 CDL-C are scaled to achieve desired angular spreads (AS) as
\begin{equation}
    \varphi_{n,scaled} = wrap\Bigl( \frac{AS_{desired}}{AS}. wrap\left(\varphi_{n}-\mu_{\varphi}\right)   + \mu_{\varphi}\Bigr).
    \label{e.AngleScaling}
\end{equation}
Per-cluster spreads are scaled as $
    c_{scaled} = ({AS_{desired}}/{AS}) \,c.$
Then, the angles of arrival/departure for rays are obtained as
\begin{equation}
    \varphi_{n,m,scaled} = \varphi_{n,scaled}+ c_{scaled} \ \alpha_m.
    \label{e.subpathAnglesScaled}
\end{equation}
Here, $AS$ and $\mu_{\varphi}$ denote the angular spreads and the power weighted mean angles of the CDL-C 901 model, calculated based on the formulas provided in \cite[Annex A]{38901-Rel19}. Also, $c$ denotes the per-cluster spreads of arrival and departure angles given in \cite[Table 7.7.1-3]{38901-Rel19}, and $\alpha_m$ represents the offset angle for ray $m$ specified in \cite[Table 7.5-3]{38901-Rel19}. Finally, we define
\begin{equation}
    wrap(x) = (x+180) \! \! \! \! \mod \!360 - 180
\end{equation}
The desired spreads $AS_{desired}$ are provided in \cite[Table 5.1.3.1-1]{38753-Rel19}\footnote{The method described here corresponds to TR 38.753 V19.0.0.}. The parameters of the rCDL-C provided in \cite[Table 5.1.3.3.1-1]{38753-Rel19}, account for the desired AS. {The intent behind this modification is to align the angular spreads with measurement campaigns conducted by the RAN4 working group. Both TR 38.901 and TR 38.753 angular spreads were derived from 5G NR measurement campaigns under different propagation conditions, and channel realizations consistent with either set may be encountered in practice. Beyond measurement alignment, this procedure provides a means of controlling the spatial correlation properties of the channel, since larger angular spreads reduce MIMO spatial correlation while smaller spreads increase it.}\\

\noindent \subsubsection{Antenna Array Virtualization:} \label{s.AAV}The antenna panel consists of single- or dual-polarized radiating elements arranged on a two-dimensional grid defined by rows and columns. The panel can be partitioned into subarrays, where each subarray comprises a set of co-polarized elements that are mapped to a single port through an antenna array virtualizer (AAV). {Using the rCDL model, and unless a steering direction is specified, broadside emission is assumed; accordingly, equal weights are applied when combining the channel coefficients of the antenna elements (in \eqref{e.CDLModel}) within each subarray. This fixed configuration enables reproducible evaluation across different equipment, decoupling the channel model from implementation-specific choices.}\\

\noindent  \subsubsection{Cluster Reduction:}
\label{s.DelayScaling} {Current RAN4 conformance test hardware supports real-time generation of channel realizations with at most 12 delay paths. Upgrading to $N=24$ delay paths, as specified in TR 38.901 CDL, would require hardware modifications. The rCDL is therefore truncated to $N'=12$ clusters to preserve compatibility with existing laboratory equipment. To minimize the impact on channel statistics, truncation retains the clusters with the highest power, while accounting for the antenna patterns and the AAV framework defined above.} To maintain the desired delay spread (DS), rescaling is applied as follows
\begin{equation}
    \tau_{n,scaled} =  \frac{DS_{desired}}{DS_{truncated}} \tau_{n},
\end{equation}
where the rms delay spread after truncation is calculated as
\begin{equation}
    DS_{truncated} = \sqrt{\frac{\sum_{n=1}^{N'} P_n \tau_{n}^2}{\sum_{n=1}^{N'} P_n} -\left( \frac{\sum_{n=1}^{N'} P_n \tau_{n}}{\sum_{n=1}^{N'} P_n}\right)^2},
    \label{e.DSrescaling}
\end{equation}
where $\tau_n$ and $P_n$ are the delay and power of the $n$-th CDL-C cluster, respectively. The delays listed in \cite[Table 5.1.3.3.1-1]{38753-Rel19} already account for the rescaling with $DS_{desired}=365$~ns.

{From the description of CDL models in Sec.~III-A and modifications detailed above, we provide} implementation guidelines for generating the rCDL-C channel in Algorithm~\ref{alg.alg1}.
\begin{algorithm}[htb]
\caption{channel generation procedure for the rCDL }\label{alg.alg1}
\begin{algorithmic}[1]
\vspace{0.1 cm}
\STATE Set \textbf{antenna array parameters} (define panel structure and orientation) and \textbf{network layout} (speed and direction of motion of UE, carrier frequency, and bandwidth).
\STATE Set \textbf{angular and delay spreads} according to the scenario.
\STATE Operate angular and/or delay \textbf{scaling} as described in III-C-\ref{s.AngleScaling} and III-C-\ref{s.DelayScaling}, respectively.
\STATE Define \textbf{sub-path angles} based on Eq. \eqref{e.subpathAnglesScaled}.
\STATE \textbf{Couple} angles within clusters using~\cite[Table 5.1.3.2-1]{38753-Rel19}.
\STATE Calculate the \textbf{field patterns} in the GCS as described in III-A-\ref{s.AntennaModeling}, by operating transformations in the Appendix.
\STATE Define the \textbf{polarization matrices} by using fixed phases defined in \cite[Table 5.1.3.2-2]{38753-Rel19}.
\STATE Calculate the \textbf{phase shifts} as in III-A-\ref{s.PhaseShifts}.
\STATE Calculate the \textbf{Doppler term} as in III-A-\ref{s.DopplerTerm}.
\STATE Apply \textbf{AAV}, if defined, as described in III-C-\ref{s.AAV}.
\end{algorithmic}
\label{alg1}
\end{algorithm}

\subsection{{Impact of Modifications on Channel Statistical Properties}}

{The impact of the modifications introduced to derive the rCDL from the RAN1 CDL on the channel statistical properties is discussed in this section. Four key properties are considered: the power-delay profile (PDP), the XPR, the DS, and the AS. The randomness reduction framework governs the initial condition and time evolution of the channel realization to reduce inter-implementation convergence time. This modification affects neither the multipath structure nor the propagation statistics; the PDP, XPR, DS, and AS are therefore left unchanged. Similarly, the AAV fixes an antenna configuration framework without modifying the underlying propagation model, leaving all statistical properties unaffected.}

{The angle scaling adjusts the per-cluster angles and per-cluster spreads according to~\eqref{e.AngleScaling}, so that the resulting angular statistics reflect measurement campaigns conducted by the RAN4 working group. This leads to different mean AoD, AoA, ZoD, and ZoA values between CDL-C~\cite[Table~7.7.1-3]{38901-Rel19} and rCDL-C~\cite[Table~5.1.3.3.1-1]{38753-Rel19}. The default AS values for both models are compared in Table~\ref{tab.AS}, where it can be observed that rCDL-C exhibits a lower ASD than CDL-C. A reduced ASD implies higher spatial correlation, which concentrates channel energy into fewer dominant spatial directions and consequently favors lower-rank transmission. The effect of this on spectral efficiency under low- and high-resolution CSI reporting codebooks is further discussed in Section~\ref{s.SE}.}

{The cluster reduction modification lowers model complexity to facilitate real-time channel generation on laboratory test equipment. The PDP is altered, as 12 clusters are retained from the 24 clusters of the angle-scaled CDL-C. To limit PDP distortion, the lowest-power clusters are discarded. The DS is preserved, as it is re-enforced after truncation per~\eqref{e.DSrescaling}. The angular spreads, however, are not re-enforced and are therefore slightly affected by truncation; the resulting values are reported in 
 Table~\ref{tab.AS}. In summary, the rCDL modifications leave the XPR and DS unchanged. The PDP is slightly altered due to cluster truncation, which also introduces a minor deviation in the angular spreads from their target values. The desired angular spread values are themselves intentionally modified to match measurements under specific propagation conditions.}

\begin{table}[h!]
\centering
\caption{{Angular spreads for the TR~38.901 CDL-C and TR~38.753 {r}CDL-C before (desired) and after truncation (truncated).}}
\label{tab.AS}
\begin{tabular}{lcccc}
\hline
\textbf{Model} & \textbf{ASD} [$^\circ$] & \textbf{ASA} [$^\circ$] & \textbf{ZSD} [$^\circ$] & \textbf{ZSA} [$^\circ$] \\
\hline
CDL-C   & 39.09 & 71.12 & 4.07 & 10.42 \\
rCDL-C; desired  & 25.76 & 74.11 & 4.90 & 18.21 \\
rCDL-C; truncated & 23.17 &	69.63	& 4.70 &	17.57 \\
\hline
\end{tabular}
\end{table}

\subsection{{Future work on rCDL}}
{Several directions for the further development of the rCDL model are outlined below.}
\begin{itemize}
    \item {\textit{Extension to MU-MIMO:}} The current rCDL formulation supports single-user (SU) MIMO evaluations. To evaluate multi-user (MU) effect related performance impacts, single- and multi-cell MU channel model extensions are required. To correctly model interference effects and their mitigation, the MU channel model needs to be spatially consistent, i.e.,  the large-scale environment should stay consistent for different positions of users, or at least the spatial properties shall consistently correlate with the positions of users and transmitters. The CDL model discussed in this paper can be extended to MU setups, by employing the spatial consistency procedures described in \cite[Sec.~7.6.3]{38901-Rel19}. Channel model parameter tables for additional spatially correlated users can be created using the procedure to move the initial user to another position, which may be close by to share common large-scale environmental features, or far away to decorrelate.
\item {\textit{Extension to other scenarios:} Channel model parameter tables are currently defined for the Urban Macrocell (UMa) scenario only. Extensions to additional deployment environments, including Urban Microcell (UMi), Indoor Factory (InF), and others, are required to broaden the applicability of the rCDL model.}
\item {\textit{Angular scaling:} Multiple angular spreads may need to be defined to reflect different propagation conditions, ranging from low to high spatial correlation, depending on the feature under evaluation. This is particularly important for the performance discrimination of high- vs. low-resolution feedback CSI reporting requirements.}
\item {\textit{PMI bias for performance evaluation:} The randomness reduction mechanism may introduce a bias in precoder performance evaluation by inducing systematic PMI preferences. A potential solution to mitigate this bias is the use of beam steering; its specification remains an active area of work.}
\end{itemize}

\section{Field Measurements}
Sections \ref{S.TDL} and \ref{S.CDL} introduced the TDL and CDL channel models and discussed their respective benefits and drawbacks. This section assesses how well these model classes capture the MIMO channel properties that govern spatial multiplexing in operational networks. Since multiplexing capability is commonly characterized by channel singular values or post-equalization SINR \cite{3gppR4-2402277}, we compare post-equalization SINR derived from measured MIMO channels with SINR profiles from CDL and TDL model realizations. For further details on angular spreads and multipath arrival characteristics, see \cite{3gppR4-2419338}.

\subsection{Measurement conditions}
{Channel matrices $\mathbf{H}$ were obtained from field measurements in a campus-type environment in Martlesham Heath, England~\cite{wsa23}. Measurements were taken at static receiver locations approximately $100$--$200$m from the base station, covering both LoS and nLoS conditions. Data were collected during office hours over $1$--$2$min intervals, producing several hundred frequency-domain channel samples per second. Occasional movement of people and vehicles occurred within the coverage sector, but no significant motion was present near the receiver.}

{The setup used reference signals transmitted in a $15$~MHz downlink channel at a carrier frequency of $2162.2$~MHz. The base station employed a commercial multi-band antenna array panel with four active RF transmit ports feeding cross-polarized antenna columns with $\pm45^{\circ}$ slant polarizations. The receiver used four RF ports connected to individual antenna elements mounted on a stationary vehicle. The measured complex channel coefficients therefore include the effects of the propagation environment and the RF chains at both transmitter and receiver.}


\subsection{Main observations}
\begin{figure}[b!]
    \centering
    \includegraphics[width=0.65\linewidth, trim=10 1 30 8, clip]{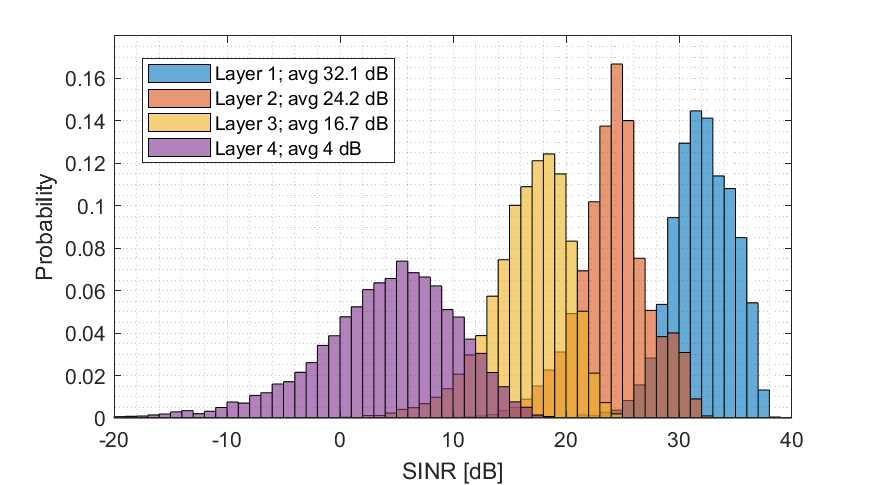}   
    \caption{Histograms of per-layer SINR assuming SVD precoding and combining based on measured MIMO channels. {Average gains of layers 2, 3 and 4 experience penalties of approximately 7, 15 and 28 dB relative to Layer 1.}}
    \label{f.btSINR}
\end{figure}

Assuming full CSI knowledge, singular value decomposition (SVD) precoding is applied at the transmitter, and the corresponding SVD combining at the receiver for each channel realization $\mathbf{H}$. This allows the typical impact of the underlying spatial MIMO propagation environment on multiplexing capability to be made explicit. The histograms in Fig.~\ref{f.btSINR} indicate a spread of approximately 28 dB between the strongest and weakest eigenmodes of the 4-by-4 MIMO channel realizations. 
Although the magnitude of the spread varies with the environment and transmission strategy, the presence of a spread is typical to practical MIMO channels.

\begin{figure}[b!]
    \centering
    \includegraphics[width=0.7\linewidth]{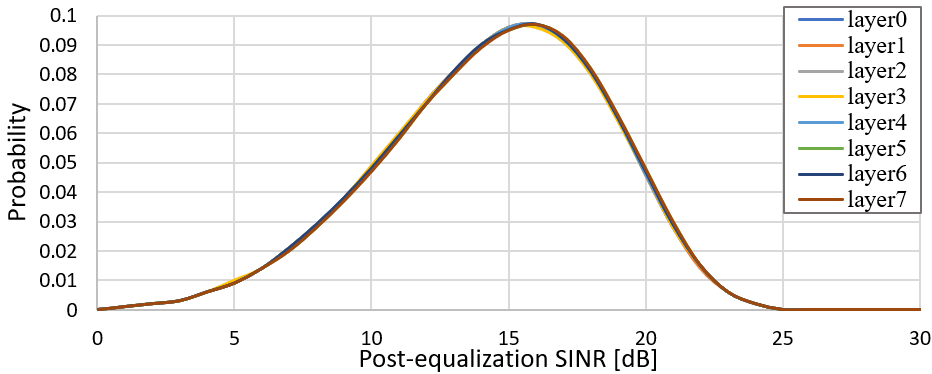}
    \caption{Simulated post-equalization SINR profile for an 8×8 MIMO system under the TDLC300 channel model {with med-A correlation}, using a randomly selected precoder from Type-I codebook and an LMMSE receiver.}
    \label{f.nokTDL}
\end{figure}
\begin{figure}[b!]
    \centering
    \includegraphics[width=0.7\linewidth]{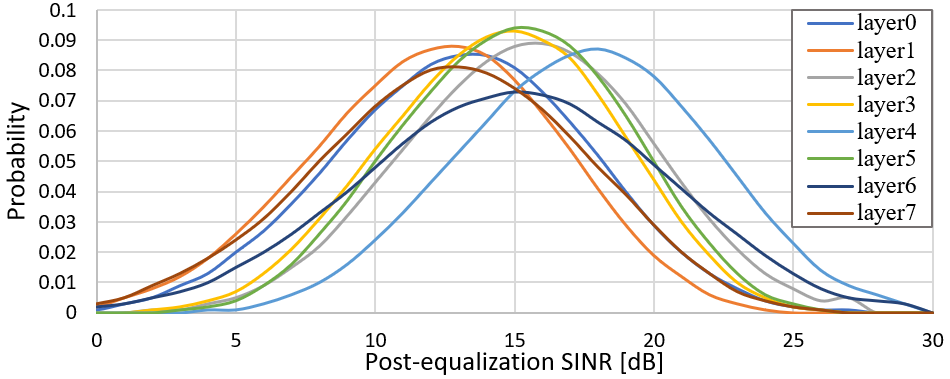}
    \caption{Simulated post-equalization SINR profile for an 8×8 MIMO system under the CDL-C channel model with 365~ns delay spread, using a randomly selected precoder from Type-I codebook and an LMMSE receiver. {A spread of $5$~dB is observed between the weakest and the strongest eigenmodes.}}
    \label{f.nokCDL}
\end{figure}

Post-equalization SINR distributions for realizations of the TDL and CDL models are provided in Fig.~\ref{f.nokTDL} and Fig.~\ref{f.nokCDL}, respectively, assuming a linear minimum mean square (LMMSE) receiver. Both figures consider an 8-by-8 MIMO system with precoding based on the 3GPP Type-I codebook. Detailed precoder description and SINR formulas are provided in Sec.~\ref{s.CSI}.  {Since the TDL model with non-low correlation concentrates the energy in a single broadside direction, random PMI corresponds to a random precoding direction for each layer with respect to the broadside direction. As a result, all layers experience the same average SINR in Fig.~\ref{f.nokTDL}.

In contrast, Fig.~\ref{f.nokCDL} {shows that applying the same transmission strategy to a CDL channel yields a spread of approximately 5~dB between the strongest and weakest eigenmodes, which is more consistent with the measured behavior.} Designing a transmission strategy that exploits the practically observed eigenmode spread is nontrivial, as it entails, for example, selecting the number of spatial layers to activate as shown in Sec.~\ref{s.PHYabstraction}, as well as determining the power allocation across the active layers. Consequently, both the development of transmission schemes and the specification of meaningful performance requirements for MIMO systems should be carried out using models that are capable of capturing this overall eigenmode-spread effect.

\section{Application to CSI Feedback based on Low- and High-Resolution Codebooks}
\label{s.CSI}
3GPP adopts a limited CSI feedback framework in which the UE measures the downlink channel based on CSI reference signals (CSI-RS), then feeds back a CSI report to the BS. The CSI report includes a rank indicator (RI), a channel quality indicator (CQI), and a precoding matrix indicator (PMI) selected from a standardized codebook \cite{38214-Rel19}. In this section, we evaluate low- and high-resolution PMI reporting using the Release 15 Type-I and Release 16 enhanced Type-II (eType-II) codebooks under legacy TDL and rCDL channel models as a case study for MIMO performance evaluation.
%
\subsection{3GPP Downlink MIMO Codebooks}
\label{s.Codebook}
To balance the trade-off between CSI resolution and feedback overhead, 3GPP employs precoder compression based on inverse DFT (IDFT) bases. We define the spatial domain (SD) compression applicable to all 3GPP codebooks specified for 5G New Radio (NR). Consider a two-dimensional planar transmit array with $N_1$ and $N_2$ antenna ports in the horizontal and vertical dimensions, respectively. SD compression selects oversampled 2D-IDFT vectors from a grid, where each vector points to a specific region (oversampling improves angular accuracy). The vectors are defined by the Kronecker product
\begin{subequations}
\begin{equation}
\mathbf{u}_{n_1,n_2}(q_1,q_2) =\mathbf{u}'_{n_1}(q_1)\otimes\mathbf{u''}_{n_2}(q_2) \in \mathbb{C}^{N_1N_2},
\end{equation}
where the horizontal and vertical IDFT vectors are defined as
\begin{equation}
\mathbf{u}'_{n_1}(q_1) = \begin{bmatrix}
 1 \!\! & \!\! e^{j\frac{2\pi n_1}{N_1}} \!\!& \!\! \hdots & \!\! e^{j\frac{ 2\pi n_1(N_1-1)}{N_1}}\!\!& 
\end{bmatrix}^T \odot  \mathbf{v}_1(q_1),
\end{equation}
\begin{equation}
\mathbf{u}''_{n_2}(q_2) = \begin{bmatrix}
   1 \!\! & \!\! e^{j\frac{2\pi n_2}{N_2}} \!\!& \!\! \hdots \!\! & \!\! e^{j\frac{ 2\pi n_2(N_2-1)}{N_2}} \!\!& 
\end{bmatrix}^T \odot  \mathbf{v}_2(q_2),
\end{equation}
where $\odot$ denotes the Hadamard product, and the rotation vector is
\begin{equation}
    \mathbf{v}_i(q_i) = \begin{bmatrix}
       \left( e^{j\frac{ 2\pi q_i}{O_iN_i}} \right)^0 \!\! & \!\!  \hdots \!\! & \!\! \left( e^{j\frac{ 2\pi q_i}{O_iN_i}} \right)^{N_i-1}
    \end{bmatrix}^T, \ \text{for }i=1,2.
    \label{e.vi}
\end{equation}
\label{e.uvector}
\end{subequations}Here, $n_1\in\{0,\ldots,N_1-1\}$, $n_2\in\{0,\ldots,N_2-1\}$, $q_1\in\{0,\ldots,O_1-1\}$, and $q_2\in\{0,\ldots,O_2-1\}$, where $O_1$ and $O_2$ denote the horizontal and vertical oversampling factors, respectively \cite{WIMOB}. This construction defines a grid $\mathcal{G}$ of $N_1O_1\times N_2O_2$ beams from which the dominant directions are selected for SD compression. \vspace{0.15 cm}

\noindent 
\subsubsection{Type-I Codebook:} For rank $1$, the precoder is given by 
\begin{equation}
  \mathbf{w} =\frac{1}{\sqrt{2N_1N_2}}\begin{bmatrix}\mathbf{u}_{n_1,n_2}(q_1,q_2)              \\
    \varphi \ \mathbf{u}_{n_1,n_2}(q_1,q_2)
  \end{bmatrix} \in \mathbb{C}^{2N_1N_2},
  \label{e.typeI_rank1_v1}
\end{equation}
where the dominant beam is selected from the grid $\mathcal{G}$, and the co-phasing between polarizations $\varphi$ is chosen from the set $\{1,j,-1,-j\}$. The precoder can equivalently be expressed as the product of an SD compression (beam) matrix $\mathbf{W}_1\in\mathbb{C}^{2N_1N_2\times2}$ and a co-phasing vector $\mathbf{w}_2\in\mathbb{C}^{2}$, that is
\begin{subequations}
\begin{align}
  \mathbf{w} & =\frac{1}{\sqrt{2N_1N_2}} \mathbf{W}_1 \mathbf{w}_{2},\\
                     & =\frac{1}{\sqrt{2N_1N_2}}
                     \begin{bmatrix}
                     \mathbf{u}_{n_1,n_2}(q_1,q_2) & \mathbf{0}_{N_1N_2\times1} \\
      \mathbf{0}_{N_1N_2\times1} & \mathbf{u}_{n_1,n_2}(q_1,q_2)
    \end{bmatrix}
  \begin{bmatrix}
      1             \\
      \varphi
    \end{bmatrix}.
\end{align}
\label{e.typeI_rank1_v2}
\end{subequations}
In \eqref{e.typeI_rank1_v1} and \eqref{e.typeI_rank1_v2}, a single co-phasing value is applied over the entire bandwidth, referred to as Type-I wideband (WB). A higher-resolution alternative assigns a precoder to each group of subcarriers, called subband (SB). The Type-I SB selects co-phasing values $\varphi_t$ (in $\mathbf{w}_{2,t}$) for each SB $t\in\{0,\ldots,N_3-1\}$, where $N_3$ denotes the number of SBs \cite{3GPPCodebooks}.
\vspace{0.15 cm}

\noindent \subsubsection{Enhanced Type-II Codebook:} Unlike Type-I, this codebook forms weighted combinations of $L\in\{2,3,4\}$ beams, yielding higher-resolution precoding that more closely approaches optimal eigen-beamforming. Due to dual polarization, the SD basis has a block-diagonal structure, that is
\begin{subequations}
\begin{equation}
\mathbf{W}_1=\begin{bmatrix}\mathbf{W}_{1p} & \mathbf{0}_{N_1N_2\times L}\\[4pt]\mathbf{0}_{N_1N_2\times L} & \mathbf{W}_{1p}\end{bmatrix}
\in\mathbb{C}^{2N_1N_2\times 2L},
\end{equation}
with
\begin{equation}
\mathbf{W}_{1p}=\begin{bmatrix}
\mathbf{u}_{n_1^{(0)},n_2^{(0)}}(q_1,q_2) & \!\!\! \cdots \!\!\! & \mathbf{u}_{n_1^{(L-1)},n_2^{(L-1)}}(q_1,q_2)
\end{bmatrix}   .
\end{equation}
\label{e.BeamMatrixTypeII}
\end{subequations}
\noindent
The IDFT vectors in \eqref{e.BeamMatrixTypeII} are selected from the grid $\mathcal{G}$ and use the same rotation indices $(q_1,q_2)$ to ensure mutual orthogonality of the SD basis vectors. 

The beams are combined using complex weights, referred to as beam combining coefficients (BCCs). A BCC is computed for each beam $i\in\{0,\ldots,L-1\}$, SB $t\in\{0,\ldots,N_3-1\}$, and spatial layer $l\in\{1,\ldots,\nu\}$, where $\nu$ denotes the transmission rank. This yields a layer-specific BCC matrix $\mathbf{W}_2^{(l)}\in\mathbb{C}^{2L\times N_3}$.

The eType-II codebook further applies frequency domain (FD) compression to exploit channel sparsity in the beam–delay domain, where each beam is associated with a limited number of delays. FD compression selects $M_\nu$ (rank-dependent) dominant delay vectors from an $N_3\times N_3$ DFT matrix (without oversampling), whose columns are
\begin{equation}
   \mathbf{y}_{n_3} = \begin{bmatrix}
   1  \!\!  & \!\! e^{j\frac{2\pi n_3}{N_3}} \!\! & \!\! \ldots \!\! & \!\! e^{j\frac{ 2\pi n_3(N_3-1)}{N_3}}
\end{bmatrix}^T, \  n_3 \in \{0,\ldots,N_3-1\}.
\label{e.yvector}
\end{equation}
The resulting layer-specific FD basis $\mathbf{W}_f^{(l)} \in \mathbb{C}^{N_3 \times M_\nu}$ for layer $l$ is applied to compress the BCC matrix as follows
\begin{equation}
    \overline{\mathbf{W}}_2^{(l)}= \mathbf{W}_2^{(l)} \mathbf{W}_f^{(l)} \in \mathbb{C}^{2L\times M_\nu},
    \label{e.BCCcompression}
\end{equation}
To further reduce the feedback overhead, only the most significant $K^{NZ}<2LM_\nu$ combining coefficients in $\overline{\mathbf{W}}_2^{(l)}$ are  quantized, yielding the reported BCC matrix $\widetilde{\mathbf{W}}_2^{(l)}$. 

At the BS, the eType-II precoder associated with layer $l$ is reconstructed as
\begin{equation}
   \mathbf{W}^{(l)} = \mathbf{W}_1 \widetilde{\mathbf{W}}_2^{(l)}\mathbf{W}_f^{(l)^H} \in \mathbb{C}^{2N_1N_2 \times N_3},
   \label{e.eTypeII}
\end{equation}
where $(.)^H$ is the conjugate transpose operator. It should be noted that normalization is applied to ensure that, for each SB $t$, the precoder $\mathbf{W}_t \in \mathbb{C}^{2N_1N_2 \times \nu}$,  satisfies $\mathbf{W}_t^H \mathbf{W}_t=\frac{1}{\nu}\mathbf{I}_\nu$, where $\mathbf{I}_\nu$ is the $\nu \times \nu$ identity matrix (Further details in \cite{3GPPCodebooks}).
\subsection{Precoder Selection Strategy}
\label{s.SelectionStrategy}
To select the best precoder from the 3GPP codebooks, we proceed as follows. First, the beam matrix is determined by selecting the rotation indices $(q_1,q_2)$ and $L$ IDFT vectors from the grid $\mathcal{G}$, with $L=1$ for Type-I and $L>1$ for eType-II. This selection can be based on the wideband channel covariance per polarization, $\mathbf{R}_{11}$ and $\mathbf{R}_{22}$, defined as
\begin{equation}
    \mathbf{R} = \begin{bmatrix}
        \mathbf{R}_{11} & \mathbf{R}_{12} \\
        \mathbf{R}_{12}^H & \mathbf{R}_{22}
    \end{bmatrix} = \frac{1}{N_3} \sum_{t=0}^{N_3-1} \mathbf{H}_t^H \mathbf{H}_t,
\end{equation}
where $\mathbf{H}_t\in\mathbb{C}^{N_R\times 2N_1N_2}$ is the representative channel for SB $t$, obtained from CSI reference signals (CSI-RS)–based channel estimation. A detailed beam selection algorithm is given in~\cite[Algorithm~1]{WIMOB}.
\vspace{0.15 cm}

\noindent \subsubsection{Type-I Precoder:}
We describe the rank-$1$ case; higher ranks follow similarly. After beam selection, the co-phasing value $\varphi\in\{1,j,-1,-j\}$ is chosen to minimize interference. For each candidate $\varphi$, SINRs are computed using the precoder $\mathbf{w}(\varphi)$ and channel estimates $\mathbf{H}_{j,t}$ for each physical resource block (PRB) $j\in\{0,\ldots,N_{SB}-1\}$ within SB $t$, where $N_{SB}$ is the number of PRBs per SB. Considering an LMMSE receiver, the SINR (for rank $1$) is given as
\begin{equation}
  \gamma_{j,t}(\varphi) = \frac{P}{\sigma_n^2} \left(\mathbf{w}(\varphi)^{H} \ \mathbf{H}_{j,t}^{H} \ \mathbf{H}_{j,t} \  \mathbf{w}(\varphi) \right),
  \label{sinr}
\end{equation}
where $P$ is the transmit power and $\sigma_n^2$ is the noise variance. The SINRs are mapped to mutual information (MI) values for modulation order $Q$ via $\mathcal{I}_Q(\cdot)$ and averaged over the bandwidth as follows
\begin{equation}
    \frac{1}{N_3} \frac{1}{N_{SB}} \sum_{t=0}^{N_3-1} \sum_{j=0}^{N_{SB}-1} \mathcal{I}_Q(\gamma_{j,t}(\varphi)).
    \label{e.avgMI}
\end{equation}
The co-phasing $\varphi$ that maximizes \eqref{e.avgMI} is selected.\vspace{0.15 cm}

\noindent \subsubsection{Enhanced Type-II Precoder:}
Let the effective channel after SD compression be
$
    \overline{\mathbf{H}}_t = \mathbf{H}_t \mathbf{W}_1 \in \mathbb{C}^{N_R \times 2L}  .
$
The combining coefficients are chosen to maximize the effective channel power under orthogonality constraints, i.e., 
\begin{align}
    & \mathbf{w}_{2,t}^{(l)} =\arg\max_{\mathbf{w}_2\in \mathbb{C}^{2L} \setminus \{\mathbf{w}_{2,t}^{(l')}\}_{l'=1}^{l}} \mathbf{w}_2^H  \overline{\mathbf{H}}_t^H \overline{\mathbf{H}}_t \mathbf{w}_2,\\
    & s.t. \quad \mathbf{w}_{2,t}^{(l)^H} \mathbf{w}_{2,t}^{(l')}=0,  
 \quad  \text{for} \ l'\neq l.
    \label{e.criterionBCC}
    \end{align}
The optimal solution for SB $t$ and layer $l$ is the eigenvector associated with the $l$-th largest eigenvalue
\begin{equation}
    \mathbf{w}_{2,t}^{(l)} = eig\Bigl(\overline{\mathbf{H}}_t^H \overline{\mathbf{H}}_t\Bigr)_l,
    \label{e.EVD}
\end{equation}
Phase normalization per-layer is then applied to $\mathbf{w}_{2,t}^{(l)}$ with respect to its largest-magnitude element, producing the BCC matrix $\widehat{\mathbf{W}}_2^{(l)}$. The FD basis for layer $l$ is subsequently formed by sequentially selecting the $M_\nu$ dominant delays according~to
\begin{equation}
   n_3^{(f,l)}= \arg\max_{\substack{
      n_3 \in \{0,\ldots,N_3-1\} \setminus \\ 
      \{n_3^{(0,l)},\ldots,n_3^{(f-1,l)}\}
   }} \left(\widehat{\mathbf{W}}_2^{(l)}  \mathbf{y}_{n_3}\right)^H   \left( \widehat{\mathbf{W}}_2^{(l)} \mathbf{y}_{n_3} \right)
\end{equation}
where $\mathbf{y}_{n_3}$ is defined in \eqref{e.yvector}. Finally, the BCC matrix is compressed according to \eqref{e.BCCcompression}, normalized, and quantized.
\subsection{Physical-layer Abstraction}
\label{s.PHYabstraction}
After PMI selection, the UE typically performs link adaptation to determine the transmission rank and the modulation and coding scheme (MCS), from which the CQI is derived. This process targets a given block error rate (BLER) based on an estimate of the effective SNR obtained from the CSI-RS channel estimate and given the PMI selected for each rank. In the RAN4 methodology, fixed MCS and rank values are assumed to facilitate comparison and cross-simulator alignment.

For performance evaluation, the SNR at the UE is estimated assuming an LMMSE receiver. The evaluation is conducted at the PRB level, i.e., based on $\mathbf{H}_{j,t}$ for PRB $j$ within SB $t$. As we consider a single-user (SU)-MIMO scenario, it is reasonable to consider that the precoder selected at the UE-side, denoted $\mathbf{W}_t^{(\nu)}$, is used for downlink data transmission on $\nu$ spatial layers. The corresponding ground-truth precoded channel is given as $\mathbf{G}_{j,t}^{(\nu)}=\mathbf{H}_{j,t} \mathbf{W}_t^{(\nu)} \in \mathbb{C}^{N_R \times \nu}$. Demodulation reference signal (DMRS) channel estimation is applied at the UE to measure the precoded channel, yielding $\widehat{\mathbf{G}}_{j,t}^{(\nu)}$. Based on this estimate, the LMMSE equalizer is given by
\begin{equation}
    \mathbf{F}_{j,t}^{(\nu)^H} = \widehat{\mathbf{G}}_{j,t}^{(\nu)^H} \left( \widehat{\mathbf{G}}_{j,t}^{(\nu)} \widehat{\mathbf{G}}_{j,t}^{(\nu)^H} + \sigma_n^2/P \ \mathbf{I}_{N_R}\right)^{-1} \in \mathbb{C}^{\nu \times N_R}. 
    \label{e.LMMSEFilter}
\end{equation}
Using the filter in \eqref{e.LMMSEFilter} and the ground-truth channel, the SINR at the output of the LMMSE receiver is estimated for each PRB.  Let $\mathbf{g}_{j,t}^{(l,\nu)}$ denote the $l$-th column of $\mathbf{G}_{j,t}^{(\nu)}$ and $\mathbf{f}_{j,t}^{(l,\nu)H}$ the corresponding row of $\mathbf{F}_{j,t}^{(\nu)H}$. The SINR for PRB $j$ within SB $t$ is then given by
\begin{subequations}
\begin{equation}
    \gamma_{j,t}^{(l,\nu)} = \frac{|\mathbf{f}_{j,t}^{(l,\nu)^H }\mathbf{g}_{j,t}^{(l,\nu)}|^2}{\mathbf{f}_{j,t}^{(l,\nu)^H} \mathbf{C}_{j,t}^{(l,\nu)} \ \mathbf{f}_{j,t}^{(l,\nu)}}
\end{equation}
with the covariance of noise-plus-interference defined as
\begin{equation}
    \mathbf{C}_{j,t}^{(l,\nu)} = \sum_{l'\neq l} \left( \mathbf{g}_{j,t}^{(l',\nu)} \mathbf{g}_{j,t}^{(l',\nu)^H} \right) + \sigma_n^2/P \ \mathbf{I}_{N_R}.
\end{equation}
\label{e.SINRPE}
\end{subequations}
The resulting SINR values per PRB and spatial layer are mapped to a single effective value over the bandwidth for the rank $\nu$ and modulation $Q$, corresponding to the fixed MCS. This mapping is performed using the mutual information effective SINR mapping (MIESM) \cite{cipriano2008MIESM} as
\begin{equation}
\gamma_{Q}^{(\nu)}=\mathcal{I}_Q^{-1}\Big( \frac{1}{\nu}  \frac{1}{N_3} \frac{1}{N_{SB}} \sum_{l=1}^\nu \sum_{t=0}^{N_3-1}\sum_{j=0}^{N_{SB}-1} \mathcal{I}_Q(\gamma_{j,t}^{(l,\nu)}) \Big)
\end{equation}
where $\mathcal{I}_Q^{-1}(\cdot)$ is the inverse of $\mathcal{I}_Q(\cdot)$. The resulting effective SINR is compared with the SINR corresponding to the target BLER for the selected MCS. This reference SINR is obtained from BLER-versus-SNR lookup tables derived from LDPC decoding. The comparison yields an effective BLER, interpreted as the retransmission probability. {Chase Combining HARQ is considered, where the receiver applies maximum ratio combining (MRC) across identical retransmitted copies. The effective SNR after $N_{rt}$ retransmissions is updated as:}
\begin{equation}
\gamma_{j,t}^{\prime(l,\nu)}   = \gamma_{j,t}^{(l,\nu)} + 10 \log_{10}(N_{rt}+1),  
\end{equation}
{where $N_{rt}<N_{rt,\max}$. This expression follows directly from the MRC combining gain under a quasi-static channel assumption, whereby the channel remains constant across all HARQ rounds. Under this assumption, the received SINR is identical at each retransmission round, and each additional retransmission contributes a 3~dB gain.} The spectral efficiency is then computed as a function of the selected MCS, the transmission rank, and the number of retransmissions.


\subsection{Simulation Results and Discussions}
\label{s.SE}
In this section, we evaluate the spectral efficiency of the Type-I (low-resolution) and eType-II (high-resolution) codebooks in a SU-MIMO scenario, as introduced in Sec.~\ref{s.Codebook}. Performance is assessed using the PMI selection strategies in Sec.~\ref{s.SelectionStrategy} together with the physical-layer abstraction framework in Sec.~\ref{s.PHYabstraction}. Both ideal conditions and practical pilot-based channel estimation are considered. In the latter case, CSI-RS and DMRS pilots are generated and mapped onto the time–frequency grid according to the standard \cite{38211-Rel18}. DMRS-based estimation averages consecutive least-squares (LS) estimates across PRBs within each frequency-domain bundle, assuming the channel is approximately constant over the bundle \cite{DAHLMAN2024189}. CSI-RS–based estimation uses a mismatched LMMSE approach assuming a flat power-delay profile for second-order statistics \cite{PIMRC}. 

The primary objective is to identify configurations in which the performance gap between eType-II and Type-I is sufficiently large to distinguish their implementations. We focus on four parameters critical for realistic evaluation of 5G and beyond systems: (1) antenna panel configuration, (2) channel model, (3) channel estimation accuracy, and (4) angular spreads. A secondary objective is to demonstrate cross-simulator alignment for the considered channel model, supporting its suitability for defining performance requirements.

\begin{table}[b!]
    \caption{Simulation parameters}
\begin{tabular}{p{7.6cm}|p{7.6cm}}
\hline
Antenna ports at BS & \((N_1,N_2)=(8,2)\) for CDL model \\
& \((N_1,N_2)=(4,2)\) for TDL model\\
\hline
Antenna ports at UE & \((N_1,N_2)=(2,1)\) \\
\hline
Antenna elements spacing & \(d=0.5 \lambda_0\) \\
\hline
Polarization slant angles at BS & \(\zeta=+/-45^\circ\) \\
\hline
Polarization slant angles at UE & \(\zeta=0/90^\circ\) \\
\hline
Antenna radiation pattern at BS & Sector as in Table 7.3-1 of \cite{38901-Rel19} \\
\hline
Antenna radiation pattern at UE & Isotropic \\
\hline
AAV & Direct port-to-element mapping \\
\hline
BS panel orientation & \((\alpha,\beta,\gamma)=(0^\circ,10^\circ,0^\circ)\) \\
\hline
UE panel orientation & \((\alpha,\beta,\gamma)=(180^\circ,0^\circ,0^\circ)\) \\
\hline
BS / UE height / BS-UE distance & 25 m / 1.5 m / 100 m \\
\hline
Carrier frequency & 3.5 GHz \\
\hline
Number of SB / PRBs per SB & \(N_3=14\) / \(N_{SB}=8\) \\
\hline
Bandwidth / Sub-carrier spacing & 40 MHz / 30 kHz \\
\hline
UE speed / travel direction & 3 km/h / \((\phi_v,\theta_v)=(65^\circ,90^\circ)\) \\
\hline
CSI/PMI reporting delay & 7 ms \\
\hline
Modulation and coding scheme & MCS 7 from Table 5.1.3.1-2 of \cite{38901-Rel19} \\
\hline
Rank/number of spatial layers & $\nu=4$ \\
\hline
Retransmission type & Chase with \(N_{RT,max}=4\) \\
\hline
Type-II codebook parameters & \textit{paramCombination-r16}=6\\
\hline
Energy per RE (EPRE) per antenna port~\cite{3gppR1-2410246} & only global power constraint considered\\
\hline
DMRS channel estimation & Type-1 with bundle size of 2 PRBs (baseline) or 4 (improved) \\
\hline
CSI-RS pilot positions & Row 17 of Table 7.4.1.5.3-1 of \cite{38211-Rel18} \\
\hline
CSI-RS to PDSCH EPRE & 1 (baseline) or 4 (improved) \\
\hline
\end{tabular}
\label{t.SimuParams}
\end{table}

Monte Carlo simulations are conducted using the legacy TDLC300 model with medium-A correlation as defined in~\cite{38101-4-Rel19}, as well as the rCDL-C model generated according to Algorithm~\ref{alg.alg1}, with an rms delay spread of 365~ns. 
{The simulation parameters listed in  Table~\ref{t.SimuParams} follow the assumptions specified in Tables 6.1-1, 6.2-1, and 6.2-2 of TR 38.753, in order to ensure consistency with the 3GPP study item on spatial channel model for demodulation performance requirements.}\\

\noindent \subsubsection{Comparison of TDL and CDL under Ideal Channel Estimation:}

This section complements the observations in Sec.~\ref{S.TDLSpatialProfile} regarding the limitations of the TDL model and shows that rCDL is suitable for evaluating CSI reporting requirements. Figs.~\ref{f.TDL} and~\ref{f.CDL_753_PE} present the normalized spectral efficiency  of Type-I and eType-II under the TDLC300 and rCDL-C models, respectively. {Normalization is applied with respect to the maximum throughput for the considered rank and MCS.} For reference, two additional curves are included in all figures. The first is a Type-I random PMI lower-bound (with uniformly drawn PMI). The second is an eigen-beamforming upper-bound, computed for each layer $l$ and SB $t$ as the $l$-th dominant eigenvector of the channel, i.e., $eig(\mathbf{H}_t^H\mathbf{H}_t)_l$, with equal power allocated across layers. This upper-bound is shown only when ideal channel estimation is considered, as it requires high estimation accuracy and is otherwise degraded.
\begin{figure}[h!]
    \centering
    \includegraphics[width=0.6\linewidth, trim = 20 0 35 10, clip]{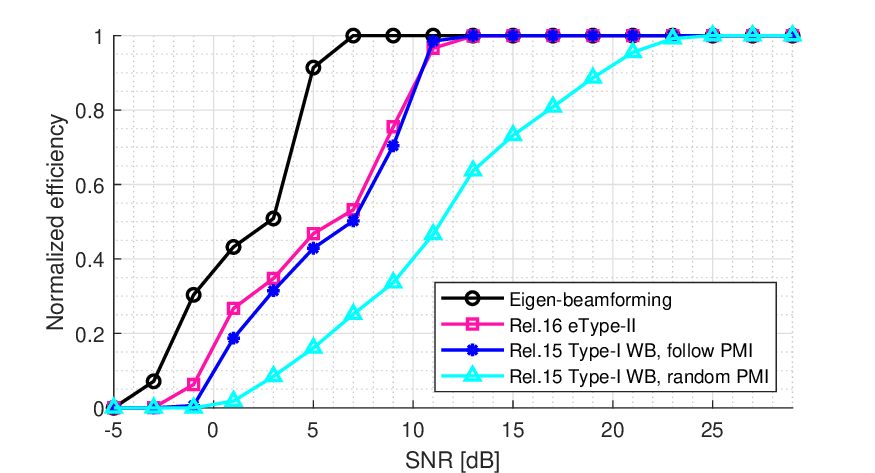}
    \caption{Throughput performance of low- and high-resolution codebooks under TDL-C300 with medium-A antenna correlation, {16 TX and 4 RX antennas,} and ideal channel estimation.}
    \label{f.TDL}
\end{figure}

{For TDL-based evaluations, we use $16$ transmit antenna ports, as the standard does not specify this model for higher antenna counts~\cite{38101-4-Rel19}.} The higher the number of antenna ports, the more degrees of freedom are available for beam selection, which can be exploited by eType-II, thereby increasing the potential performance gap between low- and high-resolution codebooks. 

Fig.~\ref{f.TDL} shows that Type-I and eType-II achieve nearly identical throughput {under TDL with $16$Tx}.  The medium-correlated TDL profile exhibits essentially a single dominant direction (cf. Fig.~\ref{f.R4-2411300Fig6_BartlettDoATDLC}), which is well captured by the single-beam structure of Type-I. In addition, both codebooks operate significantly below the eigen-beamforming bound and relatively close to the random PMI lower-bound, indicating that they are not optimized for TDL-like channels. 

{For the same 16~Tx configuration under TR~38.901 CDL-C, a clear performance gap of approximately 5~dB is observed at 50\% of the maximum throughput between eType-II and Type-I. This confirms that, regardless of the antenna port count restriction imposed by the standard, the TDL model is unsuitable for specifying CSI reporting requirements for high- and low-resolution codebooks, as both schemes yield quasi-identical performance under TDL. In the remainder of this work, $32$Tx antenna ports are adopted for CDL-based evaluations, as this configuration is more representative of massive MIMO deployments in current commercial networks.} 

Fig.~10 shows a clear performance gap under TR 38.753 rCDL-C: the eType-II consistently achieves a gain of about 2~dB over Type-I under ideal channel estimation and operates close to the eigen-beamforming bound. {However, the gain of eType-II with respect to Type-I is reduced under TR38.753 rCDL-C compared to TR 38.901 CDL-C due to the lower angular spread (detailed in 3)}. This behavior is consistent with the eType-II design, in which linear combinations of multiple IDFT vectors approximate the channel eigenvectors. Moreover, under the rCDL model, both codebooks achieve a significant gain over the random PMI bound, as their design explicitly accounts for the spatial characteristics of MIMO propagation. 
\begin{figure}[h!]
    \centering
    \includegraphics[width=0.6\linewidth,trim = 20 0 35 10, clip]{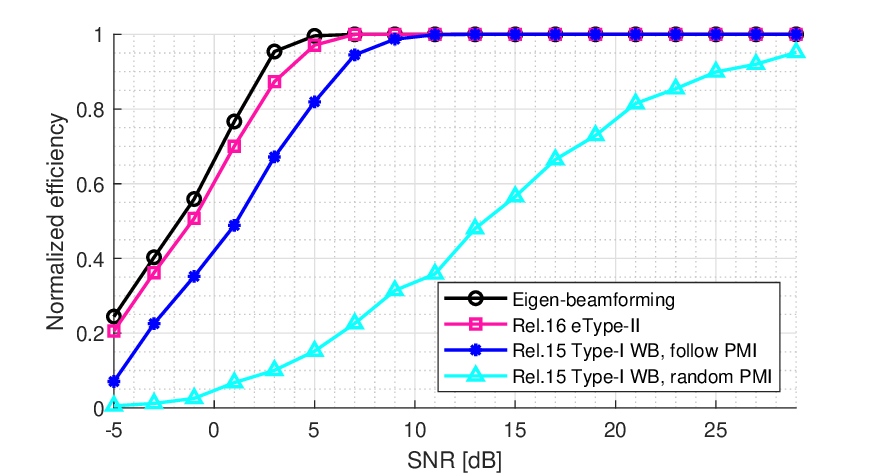}
    \caption{Throughput performance of low- and high-resolution codebooks under {TR 38.753 rCDL-C with $365$~ns delay spread, 32 TX and 4 RX antennas,} and ideal channel estimation.}
    \label{f.CDL_753_PE}
\end{figure}

{To assess the impact of the modifications made to the RAN1 CDL, we also present the spectral efficiency under TR~38.901 CDL in Fig.~\ref{f.901CDL}. To enable fair comparison, Fig.~\ref{f.753_rescaled_ideal} presents the performance under TR~38.753 rCDL-C assuming the angular spreads of TR 38.901 CDL-C. The results obtained are very similar with clear gap between Type-I and eType-II under both models. This showcases that despite modifications made (cluster truncation and randomness reduction), when considering similar angular statistics, the rCDL provides an accurate representation of MIMO propagation and approximates well the RAN1 CDL.}

\begin{figure}[h!]
    \centering
    \includegraphics[width=0.6\linewidth, trim = 20 0 35 10, clip]{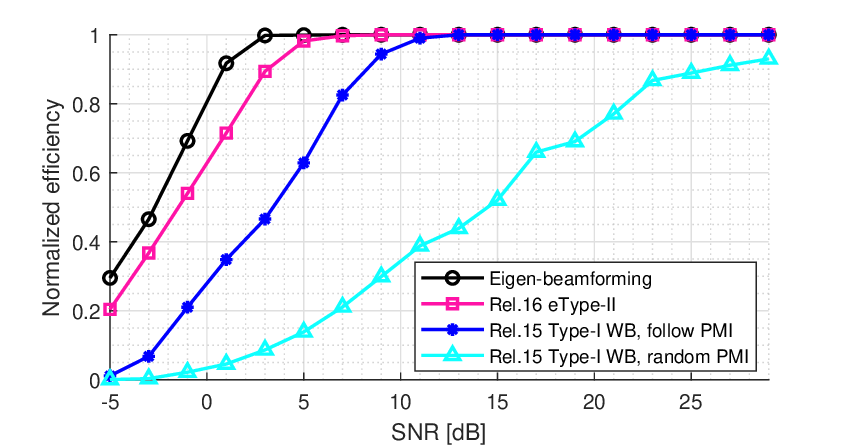}
        \caption{{Throughput performance of low- and high-resolution codebooks under  TR 38.901 CDL-C with $365$~ns delay spread, 32 TX and 4 RX antennas, and ideal channel estimation.}}
    \label{f.901CDL}
\end{figure}
\begin{figure}[h!]
    \centering
    \includegraphics[width=0.6\linewidth, trim = 20 0 35 10, clip]{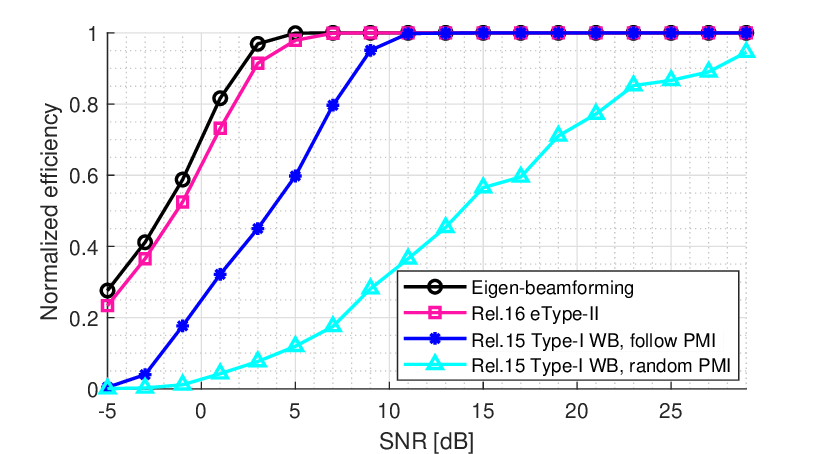}
\caption{Throughput performance of low- and high-resolution codebooks under {modified TR 38.753 rCDL-C with 365~ns delay spread, 32 TX and 4 RX antennas, and ideal channel estimation. Angular spreads of TR 38.901 CDL-C are applied using the procedure in Sec.~III-C-\ref{s.AngleScaling}.}}
    \label{f.753_rescaled_ideal}
\end{figure}

\begin{figure}[h!]
    \centering
    \includegraphics[width=0.6\linewidth, trim = 20 0 35 10, clip]{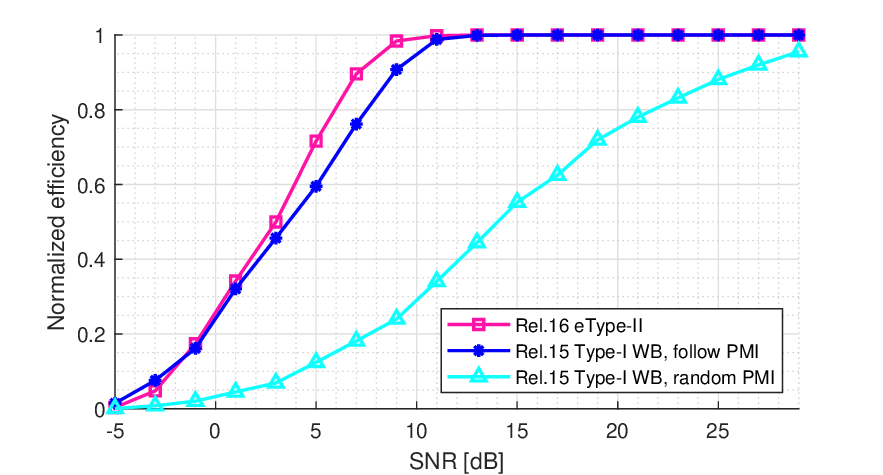}
\caption{Throughput performance of low- and high-resolution codebooks under {TR 38.753 rCDL-C with 365~ns delay spread, 32 TX and 4 RX antennas, CSI-RS channel estimation with density $\rho=1$ and unit CSI-RS-to-PDSCH EPRE ratio, and DMRS channel estimation with a bundle size of 2 PRBs.}}
    \label{f.CDL_753_IE}
\end{figure}
\noindent \subsubsection{Effect of the Channel Estimation Accuracy:} 

Fig.~\ref{f.CDL_753_IE} shows the spectral efficiency under the rCDL-C model with practical CSI-RS and DMRS channel estimation based on RAN4 settings. When comparing with Fig.~\ref{f.CDL_753_PE}, we observe that the performance gap between eType-II and Type-I decreases in the presence of estimation errors, particularly at low SNR. At 70\% of the maximum throughput, eType-II provides about 1.5~dB gain, which drops below 0.5~dB at 50\% throughput. This gap reduction occurs because high-resolution codebooks are more sensitive to estimation noise, whereas the coarser Type-I codebook is more robust by design. It should also be noted that the lower-bound is almost unchanged compared to the ideal case, as it is not affected by CSI-RS estimation errors, since there is no PMI selection.

The baseline estimation assumes equal CSI-RS and PDSCH EPRE. In practice, CSI-RS power can be boosted in the frequency domain by reallocating power from REs that are set to zero in the OFDM grid. In the following, considering pilots density and location, we apply a CSI-RS to PDSCH EPRE ratio of 4 to improve the estimation accuracy. Additionally, the DMRS bundle size is increased from $2$ to $4$ PRBs, thereby reducing estimation noise through averaging while assuming that the channel remains sufficiently stable across the bundle. Fig.~\ref{f.CDL_753_IE_improved} shows that under these improved conditions, eType-II consistently demonstrates a gain over Type-I even at negative SNR values, with the gap increasing up to 2~dB at 70\% throughput, although it remains small at low SNR.\\

\noindent\subsubsection{Effect of the Angular Spreads:}

Fig.~\ref{f.CDL_753_IE_improved_AS_901} presents performance under improved channel estimation conditions while adopting the angular spreads defined in TR~38.901 for the RAN1 CDL-C model. We observe that the performance gap increases, reaching 5~dB at 70\% throughput. This confirms that spatial diversity, reflected in the angular spreads, is a key factor for differentiating the two schemes. {This is attributable to the larger ASD in TR~38.901, which increases directional diversity and produces a richer spatial channel structure. In such conditions, the low-resolution Type-I codebook is unable to accurately capture the spatial structure, whereas the eType-II codebook retains its effectiveness.}
\begin{figure}[h!]
    \centering
    \includegraphics[width=0.6\linewidth, trim = 20 0 35 10, clip]{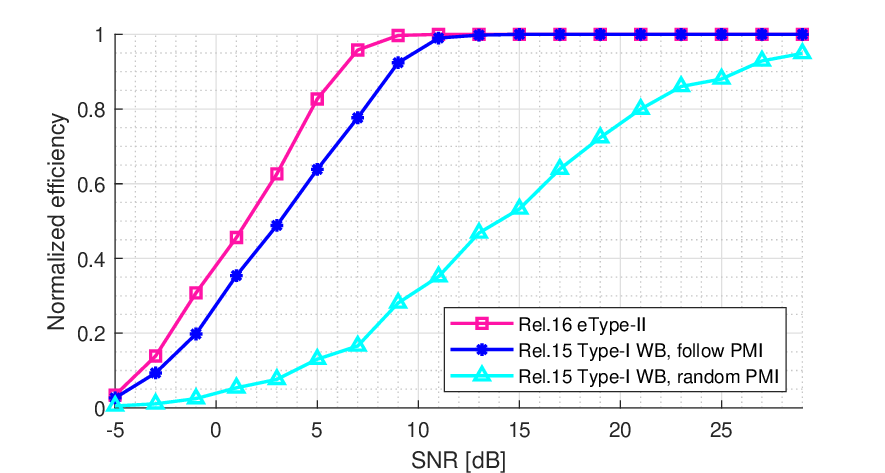}
\caption{Throughput performance of low- and high-resolution codebooks under {TR 38.753 rCDL-C with 365~ns delay spread, 32 TX and 4 RX antennas, CSI-RS channel estimation with density $\rho=1$ and CSI-RS-to-PDSCH EPRE ratio of 4, and DMRS channel estimation with a bundle size of 4 PRBs.}}
    \label{f.CDL_753_IE_improved}
\end{figure}
\begin{figure}[h!]
    \centering
    \includegraphics[width=0.6\linewidth, trim = 20 0 35 10, clip]{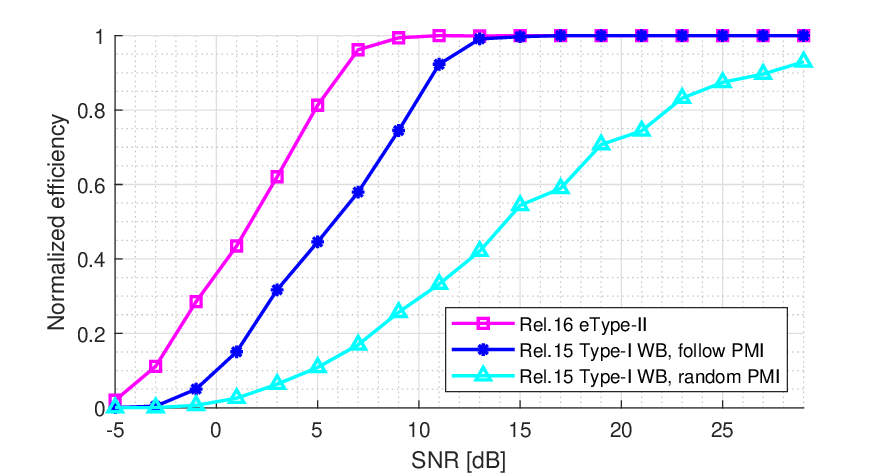}
\caption{Throughput performance of low- and high-resolution codebooks under {modified TR 38.753 rCDL-C with 365~ns delay spread, 32 TX and 4 RX antennas, CSI-RS channel estimation with density $\rho=1$ and CSI-RS-to-PDSCH EPRE ratio of 4, and DMRS channel estimation with a bundle size of 4 PRBs. Angular spreads of TR 38.901 CDL-C are applied using the procedure described in Sec.~III-C-\ref{s.AngleScaling}.}}
    \label{f.CDL_753_IE_improved_AS_901}
\end{figure}

\noindent\noindent\subsubsection{Cross-simulator Alignment:} {A comparison of results obtained from implementations by multiple companies participating in the study item was performed under the rCDL channel model with default angular and delay spreads as specified in~\cite[Table~5.1.3.3.1-1]{38753-Rel19}. Simulation parameters follow those listed in Table~\ref{t.SimuParams}, except that 8 Tx arranged as $(N_1,N_2 )=(4,1)$ and 4 Tx arranged as $(N_1,N_2 )=(2,1)$ are considered rather than 32 Tx, in accordance with Tables 6.1-1, 6.2-1, and 6.2-2 of TR 38.753.} The SNR values corresponding to 70\% and 90\% of the maximum throughput for 4-layer transmission, as reported by multiple companies in Table 7.1-3 of~\cite{38753-Rel19}, exhibit a spread of less than 2.5~dB, {where our results correspond to source 9 in that table.} This indicates a satisfactory degree of cross-simulator alignment for the rCDL-C model, supporting its suitability for defining MIMO performance requirements.

\section{Conclusion}
This paper investigated the rCDL model introduced in the RAN4 SI on channel modeling for performance requirements. We reviewed the legacy TDL model currently used by RAN4 and then described the derivation of the rCDL from the RAN1 CDL model specified in TR 38.901. The reduced model is designed to preserve the main propagation characteristics {while reducing inter-implementation convergence time, making it suitable for real-time RAN4 conformance testing.}

A comprehensive comparison between the rCDL and legacy TDL models was conducted using multiple approaches. First, spatial profile analysis showed that the rCDL captures multiple stable angular directions, whereas the TDL model is spatially agnostic. Second, measurements conducted in a typical MIMO deployment confirmed that CDL-based modeling more accurately reflects practical propagation conditions. Finally, as a case study, CSI reporting for SU-MIMO was evaluated. The results showed that the rCDL model enables clear discrimination between high- and low-resolution codebooks, whereas legacy TDL models do not. Hence, despite the simplifications introduced relative to the RAN1 CDL, the rCDL still yields meaningful MIMO performance assessment. We also identified parameters that require careful configuration to ensure valid evaluation of CSI feedback techniques. {Future work includes extending the rCDL to MU-MIMO scenarios using the spatial consistency procedures of TR 38.901, and broadening the model applicability to additional deployment scenarios such as Urban Microcell (UMi) and Indoor Factory (InF). Multiple angular spread configurations should also be defined to reflect varying propagation conditions depending on the feature under evaluation. Finally, the PMI bias introduced by the randomness reduction mechanism requires further investigation; beam steering is a candidate solution.}

\section*{Acknowledgments}
The authors are deeply grateful to all the colleagues who generously shared their time and experiences for this research. In particular, we want to thank Yann Léost from Nokia Networks France and Amol Dhere from Nokia Denmark for their contributed link layer expertise and results. Their contributions have been instrumental in the success of this work.

\begin{appendices}

\section{Transformation of the coordinate systems}
\label{A.Transfo}
\noindent\textit{Definition: Transformation between the GCS and LCS}\\
Let the coordinates of a point be $\hat{\mathbf{\rho}} = (x,y,z)$ in the GCS and $\hat{\mathbf{\rho}}'=(x',y',z')$ in the LCS$'$, these two sets of coordinates are related via the composite rotation matrix $ \mathbf{R}$ as 
\begin{equation}
     \hat{\mathbf{\rho}} = \mathbf{R} \ \hat{\mathbf{\rho}}'
     \text{ and } 
     \hat{\mathbf{\rho}}' = \mathbf{R}^{-1} \ \hat{\mathbf{\rho}},
\end{equation}
where $\mathbf{R}^{-1}=\mathbf{R}^T$ (as $\mathbf{R}$ is orthogonal) is the reverse transformation. This rotation results from the product of the three elementary rotations described in III-A-2, that is
\begin{equation}
    \mathbf{R} = \mathbf{R}_z(\alpha) \mathbf{R}_y(\beta) \mathbf{R}_x(\gamma)  = \begin{bmatrix}
        \cos \alpha  \! \! \! & -\sin \alpha & \! \! \! 0 \\
        \sin \alpha \! \! \! & \cos \alpha & \! \! \! 0 \\
        0 \! \! \! & 0 & \! \! \! 1
    \end{bmatrix} \begin{bmatrix}
        \cos \beta \! \! \!  & 0 & \! \! \! \sin \beta \\
        0 \! \! \! & 1 & \! \! \! 0  \\
        -\sin \beta \! \! \! & 0 & \! \! \! \cos \beta
    \end{bmatrix} 
    \begin{bmatrix}
        1 \! \! \!& 0 &\! \! \! 0 \\
        0 \! \! \!& \cos \gamma &\! \! \! -\sin \gamma \\
        0 \! \! \! & \sin \gamma & \! \! \! \cos \gamma 
    \end{bmatrix}
    \label{e.rotation}
\end{equation}
The direct transformation between the GCS and LCS$''$ can be operated by considering $\mathbf{R}_x(\gamma+\zeta)$ instead of $\mathbf{R}_x(\gamma)$ in \eqref{e.rotation}.\\

\noindent \subsubsection{Angles transformation (GCS $\rightarrow$ LCS$''$)}\\
The fields are defined in \eqref{e.FieldLCSpp} based on the clusters' angles in LCS$''$. Transformation of the angles to LCS$''$ is operated using the inverse rotation matrix $\mathbf{R}^{-1}$, which yields
\begin{subequations}
\begin{equation}
    \theta'' = \arccos \left[ cos \beta \cos{(\gamma+\zeta)} \cos\theta + \sin \theta \left(\sin \beta \cos{(\gamma+\zeta)} 
 \right. \right. \left. \left.\cos{(\phi-\alpha)} 
 - \sin{(\gamma+\zeta )} \sin{(\phi-\alpha )} \right) \right]
\end{equation}
\begin{multline}
    \phi'' = \arg \left\{ \cos{\beta} \sin{\theta} \cos{(\phi-\alpha)} - \sin{\beta} \cos{\theta} + j \left( \cos{\beta} \sin{(\gamma+\zeta)} \cos{\theta}  \right. \right.\\ \left. \left. + \sin{\theta} \left[ \sin{\beta} \sin{(\gamma+\zeta)} \cos{(\phi-\alpha)} + \cos{(\gamma+\zeta)} \sin{(\phi-\alpha)}\right] \right)\right\}
\end{multline}
\label{e.anglesLCSpp}
\end{subequations}
\noindent\textit{Proof.} Consider a point $(x,y,z)$ on the unit sphere defined by the spherical coordinates $(\rho,\theta, \phi)$ (with $\rho=1$). The cartesian representation of that point is given by 
\begin{equation}
   \hat{\mathbf{\rho}} = \begin{bmatrix}
       x \\ y \\ z 
   \end{bmatrix} = \begin{bmatrix}
        \sin \theta \cos \phi  \\
        \sin \theta \sin \phi  \\
        \cos \theta
    \end{bmatrix}
    \label{e.rho}
\end{equation}
Given \eqref{e.rho} and considering that a point represented by $\hat{\mathbf{\rho}}$ in the GCS corresponds to $\mathbf{R}^{-1}\hat{\mathbf{\rho}}$ in the LCS$''$, we can show that
\begin{subequations}
    \begin{equation}
        \theta''= \arccos{(\begin{bmatrix}
            0 & 0 & 1
        \end{bmatrix} 
        \mathbf{R}^{-1} \hat{\mathbf{\rho}})},
    \end{equation}
    \begin{equation}
        \phi''= \arg{(\begin{bmatrix}
            1 & j & 0
        \end{bmatrix} \mathbf{R}^{-1} \hat{\mathbf{\rho}})},
    \end{equation}
\end{subequations}
which yields formulas in \eqref{e.anglesLCSpp}.\\

\noindent\subsubsection{Fields transformation (LCS$''$ $\rightarrow$ GCS)}\\
Double-primed fields as defined in \eqref{e.FieldLCSpp} are transformed to GCS as follows
\begin{equation}
    \mathbf{F}(\theta,\phi) = \begin{bmatrix}
        \hat{\theta}^T \mathbf{R} \hat{\theta}'' & \hat{\theta}^T \mathbf{R} \hat{\phi}'' \\
         \hat{\phi}^T \mathbf{R} \hat{\theta}'' & \hat{\phi}^T \mathbf{R} \hat{\phi}''
    \end{bmatrix} \mathbf{F}''(\theta'',\phi'')
    \label{e.FieldTransformation}
\end{equation}
It can be shown that this is equivalent to an angular displacement of $\psi$ between the two pairs of unit vectors, i.e.,
\begin{equation}
\begin{bmatrix}
        \hat{\theta}^T \mathbf{R} \hat{\theta}'' & \hat{\theta}^T \mathbf{R} \hat{\phi}'' \\
         \hat{\phi}^T \mathbf{R} \hat{\theta}'' & \hat{\phi}^T \mathbf{R} \hat{\phi}''
    \end{bmatrix} = \begin{bmatrix}
        \cos \psi & -\sin{(\psi)} \\
        \sin{(\psi)} &  \cos \psi
    \end{bmatrix}
\end{equation}
with $\psi = \arg\left\{  \hat{\theta}^T \mathbf{R} \hat{\theta}'' + j \ \hat{\phi}^T \mathbf{R} \hat{\theta}'' \right\}$.
The angle can be expressed as function of the rotation angles $(\alpha,\beta,\gamma+\zeta)$ and the spherical position $(\theta,\phi)$ as
\begin{multline}
    \psi = \arg \bigl\{  \sin{(\gamma+\zeta)} \cos \theta \sin (\phi - \alpha) + \cos{(\gamma+\zeta)} (\cos \beta \sin \theta \\- \sin \beta \cos \theta    \cos (\phi - \alpha))  + j \left[ \sin{(\gamma+\zeta)} \cos (\phi - \alpha) + \sin \beta \right. \\ \left. \cos{(\gamma+\zeta)} \sin (\phi - \alpha) \right] \bigr\}
    \label{e.psiDef}
\end{multline}
The operations described above to calculate the field in the GCS are referred to as polarization Model-1 in the standard. An alternative model (Model-2) is also defined in \cite{38901-Rel19}. The latter considers that the radiation power is split into the co-polar and cross-polar components as
\begin{equation}
    \mathbf{F}'(\theta',\phi' ) = \begin{bmatrix}
        F'_{\theta'}(\theta',\phi')\\
        F'_{\phi'}(\theta',\phi')
    \end{bmatrix} = \sqrt{A'(\theta', \phi')} \begin{bmatrix}
        \cos{\zeta}\\
        \sin{\zeta}
    \end{bmatrix}
\end{equation}
where $A'(\theta',\phi')=A''(\theta'',\phi'')$ with $\theta''=\theta'$ and $\phi''=\phi'$. The fields are then transformed from the LCS$'$ to the GCS based on \eqref{e.FieldTransformation} with $\psi$ defined as in \eqref{e.psiDef} by replacing $(\gamma+\zeta)$ by $\gamma$, as polarization is already considered.
\end{appendices}

\bibliographystyle{IEEEtran}
\bibliography{bibliography}


\vfill


\end{document}